%% file: main.tex
\documentclass[lettersize,journal]{IEEEtran}

\input{preamble}

\input{acronyms}

\begin{document}
\bstctlcite{IEEEexample:BSTcontrol}
\title{Enabling Efficient Hybrid Systolic Computation\\in Shared-L1-Memory Manycore Clusters}

\author{Sergio~Mazzola, Samuel~Riedel, and~Luca~Benini
  \thanks{Manuscript received 2 February 2024; revised 24 April 2024.}
  \thanks{This paper extends our preliminary work published in \cite{riedel2023mms}.}
  \thanks{This work was partly funded by the ETH Future Computing Laboratory (EFCL), financed by a donation from Huawei Technologies, and by the COREnext project, supported by the EU Horizon Europe research and innovation programme under grant agreement No. 101092598. Acknowledgments also go to Gua Hao Khov, Matheus Cavalcante, Vaibhav Krishna, and Marco Bertuletti for their technical support.}
  \IEEEcompsocitemizethanks{%
    \IEEEcompsocthanksitem{} Sergio~Mazzola and Samuel~Riedel are with the Integrated Systems Laboratory (IIS), Swiss Federal Institute of
     Technology (ETH Zürich), 8092 Zürich, Switzerland. E-mail:
     \{smazzola, sriedel\}@iis.ee.ethz.ch
    \IEEEcompsocthanksitem{} Luca Benini is with the Integrated Systems Laboratory (IIS), Swiss Federal Institute of
     Technology (ETH Zürich), 8092 Zürich, Switzerland, and also with the Department of Electrical, Electronic and
     Information Engineering (DEI), University of Bologna, 40126 Bologna,
     Italy. E-mail: lbenini@iis.ee.ethz.ch.
  }
}

\markboth{IEEE Transactions on Very Large Scale Integration (VLSI) Systems,~Vol.~XX, No.~X, MMMM~2024}%
{Mazzola \MakeLowercase{\textit{et al.}}: Enabling Efficient Hybrid Systolic Computation in Shared-L1-Memory Manycore Clusters}

\IEEEpubid{0000--0000/00\$00.00~\copyright~2021 IEEE}

\maketitle

\begin{abstract}
    \glsresetall
    Systolic arrays and shared-L1-memory manycore clusters are commonly used architectural paradigms that offer different trade-offs to accelerate parallel workloads.
    While the first excel with regular dataflow at the cost of rigid architectures and complex programming models, the second are versatile and easy to program but require explicit dataflow management and synchronization.
    This work aims at enabling efficient systolic execution on shared-L1-memory manycore clusters. We devise a flexible architecture where small and energy-efficient \mbox{\riscv{}} cores act as the systolic array's \glspl{pe} and can form diverse, reconfigurable systolic topologies through queues mapped in the cluster's shared memory.
    We introduce two low-overhead \mbox{\riscv{}} \acrshort{isa} extensions for efficient systolic execution, namely \emph{Xqueue} and \emph{\Glspl{qlr}}, which support queue management in hardware. The Xqueue extension enables single-instruction access to shared-memory-mapped queues, while \glspl{qlr} allow implicit and autonomous access to them, relieving the cores of explicit communication instructions.
    We demonstrate Xqueue and \glspl{qlr} in \mempool{}, an open-source shared-memory cluster with 256 \glspl{pe}, and analyze the hybrid systolic-shared-memory architecture's trade-offs on several \acrshort{dsp} kernels with diverse arithmetic intensity.
    For an area increase of just \SI{6}{\percent}, our hybrid architecture can double \mempool{}'s compute unit utilization, reaching up to \SI{73}{\percent}.
    In typical conditions (TT/\SI{0.80}{\volt}\kern-.16em/\SI{25}{\celsius}), in a \SI{22}{\nano\meter} FDX technology, our hybrid architecture runs at \SI{600}{\mega\hertz} with no frequency degradation and is up to \SI{65}{\percent} more energy efficient than the shared-memory baseline, achieving up to \SI{208}{\giga\ops\per\watt}, with up to \SI{63}{\percent} of power spent in the \glspl{pe}.
\end{abstract}


\glsresetall


\section{Introduction}\label{sec:introduction}

\IEEEPARstart{H}{igh} performance at low energy is essential for computationally demanding workloads like machine learning, computational photography, and telecommunications~\cite{hennessy2019new,muralidhar2022energy}.
Leveraging the compute-intensive, data-oblivious, and parallel nature of these workloads, massively scaled-up and heterogeneous architectures have thrived since the breakdown of Dennard scaling~\cite{taylor2012dark}.
Scaling up a parallel architecture involves a trade-off with general-purpose capabilities and programmability on one hand, and performance and energy efficiency on the other.
\textit{General-purpose processors} trade a lower energy efficiency for flexibility and programmability. Architectures such as Intel's Core-i9~\cite{IntelCorporation2023} feature large, powerful cores with complex functionalities to ease programming, which nevertheless impose a high power overhead and scalability limits.
This paper focuses on the \textit{multi-core cluster with shared L1 \gls{tcdm}} architectural pattern, extensively used in \glspl{gpu}~\cite{NVIDIACorporation2022} and embedded \glspl{soc}~\cite{GreenWavesTechnologies2021,Ginosar2016}. The success of this architecture is due to its effective trade-off between flexibility and efficiency, achieved by integrating small, agile cores sharing a low-latency, multi-banked L1 memory. This streamlines workload distribution and communication among the cores without expensive cache coherency protocols.

\IEEEpubidadjcol

However, scaling up the number of cores requires careful consideration for maintaining low-latency access to L1 memory and high instruction stream bandwidth.
GPU's \glspl{sm} address this with a \gls{simt} paradigm, sacrificing instruction stream diversity.
More flexible shared-L1 clusters can still scale up to a few hundred independent cores thanks to hierarchical interconnects~\cite{riedel2023mempool} or \glspl{noc}~\cite{olofsson2016epiphany}.
Such a memory model makes these clusters suitable for a wide range of workloads, including irregular and non-data-oblivious algorithms.
Nevertheless, many key computational kernels in machine learning and communications have a regular dataflow structure among fundamental operators (e.g., multiply-accumulate), which maps efficiently on systolic arrays.
\textit{Systolic array architectures} are regular, neighborhood-based networks of \glspl{pe}~\cite{Kung82}. Processing occurs in lockstep, facilitating implicit communication and synchronization for improved resource utilization, performance, and energy efficiency.
However, the efficiency of systolic arrays is strongly tied to their application-specific topology~\cite{jouppi2017datacenter,Redgrave2018}, which limits their usage for non-regular algorithms and complex applications with multiple heterogeneous computational kernels~\cite{sun2023sense}.

\begin{figure}
    \centering
    \includegraphics{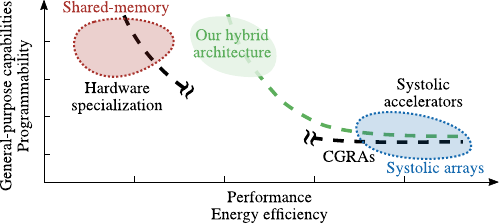}
    \caption{Qualitative Pareto front of the trade-off among flexibility, performance, energy efficiency, and programmability for massively parallel architectures. The two considered architectural templates are depicted with red and blue on the curve, along with the design space they covered throughout their evolution.}
    \label{fig:pareto}
\end{figure}

As represented in \Cref{fig:pareto}, systolic arrays and shared-L1-memory clusters occupy opposing sides in the trade-off among flexibility, programmability, performance, and energy efficiency. 
Yet, systolic architectures have been evolving towards \glspl{cgra}, broadening their applicability, and were integrated into general-purpose systems as accelerators~\cite{chen2016eyeriss,Podobas2020,fornt2023energy}. 
Simultaneously, shared-L1-memory clusters have been embracing specialized hardware, transitioning more and more to heterogeneous architectures~\cite{hennessy2019new,fuchs2019accelerator}.
Despite this convergence, the two paradigms remain separated by a clear gap.
In this work, we bridge the gap by proposing an architectural hybridization of a shared-L1-memory manycore cluster that efficiently supports peer-to-peer core communication. This enables flexible and efficient execution of systolic workloads on shared-memory, multi-core architectures without compromising their general-purpose capabilities, performance, and programmability.

Specifically, we employ user-programmable FIFO queues mapped in the cluster's shared memory to enable communication between any two cores, exploiting their low-latency access to L1. The memory-mapped queues allow for the dynamic arrangement of cores in various systolic network topologies while simultaneously supporting the concurrent, shared-memory execution model, which provides an efficient implementation of \emph{multicast} and \emph{gather} collective communication operations~\cite{Kung82,vandeGeijn2011}.
To boost systolic communication efficiency, we propose two low-overhead \riscv{} \gls{isa} extensions accelerating queue access, namely \textit{\xqueue{}} and \textit{\Glspl{qlr}}.
The \xqueue{} extension implements a parametrizable number of hardware-accelerated, memory-mapped queues with single-instruction access. At the same time, the \qlr{} extension empowers a parametrizable number of \xqueue{}-managed queues with autonomous access, which relieves the cores of explicit communication instructions.

Combining systolic communication capabilities and shared-memory flexibility unveils a software design space of unprecedented trade-offs. 
In our hybrid architecture, the software stack can benefit from both regular systolic dataflow and global, concurrent communication. We leverage the combination of the two execution models to optimize data reuse and memory accesses in many \gls{dsp} kernels, targeting improved performance and energy efficiency.
To demonstrate the proposed architecture, we enhance the open-source, \riscv{}-based \mempool{} architecture~\cite{riedel2023mempool}, a \SIadj{32}{\bit} general-purpose manycore system with 256 cores sharing a low-latency L1 \gls{spm}.
Specifically, this paper extends our preliminary work~\cite{riedel2023mms} and presents the following contributions:

\begin{itemize}
  \item We propose a new hybrid systolic-shared-memory architecture that simultaneously supports systolic dataflow and shared-memory operations, with minimal impact on \gls{ppa} figures and no influence on the system's programmability and flexibility.
  To the best of our knowledge, this is the first architecture enhancing a general-purpose manycore cluster with efficient support of systolic computation.
  \item We introduce two lightweight \riscv{} \gls{isa} extensions accelerating the cores' access to memory-mapped queues, that significantly increase the performance and energy efficiency of the systolic communication in the shared-memory cluster.
  \item We explore the hybrid systolic-shared-memory execution models enabled by the combination of systolic dataflow and global communication, analyzing the involved trade-offs. We introduce hybrid implementations of key computational kernels with high arithmetic intensity, namely matrix multiplication, 2D convolution, and \gls{fft}. \rebuttal{Furthermore, we discuss the applicability of the hybrid architecture to kernels with low arithmetic intensity, such as dot product and axpy.}
  \item We demonstrate a full implementation of the novel hybrid architecture on the \mempool{} shared-L1-memory cluster, including a fully open-source hardware architecture and software stack~\cite{MemPoolRepo}. We evaluate the impact on \gls{ppa} figures and the trade-offs for the hybrid execution model.
\end{itemize}

\noindent For a minimal area increase of \SI{6}{\percent} and no frequency degradation, the hybrid architecture reaches a utilization of up to \SI{73}{\percent} with high-arithmetic-intensity kernels, improving up to \SI{86}{\percent} with respect to the shared-memory baseline.
When implemented in GlobalFoundries' 22FDX \gls{fdsoi} technology, our hybrid architecture achieves up to \SI{208}{\giga\ops\per\watt} for 32-bit operations, running at \SI{600}{\mega\hertz} under typical conditions (TT/\SI{0.80}{\volt}\kern-.16em/\SI{25}{\celsius}). The hybrid execution model spends up to \SI{63}{\percent} of the total power in the \glspl{pe}, making the hybrid architecture up to \SI{65}{\percent} more energy efficient than the baseline.

The remainder of this paper is structured as follows. \Cref{sec:relwork} gives an overview of the related works in the literature.
\Cref{sec:architecture} presents our hybrid systolic-shared-memory architecture, along with its \gls{isa} extensions, while \Cref{sec:implementation} discusses its implementation in \mempool{}. \Cref{sec:kernels} explores the hybrid execution models for our computational kernels. Finally, \Cref{sec:evaluation} evaluates the \gls{ppa} of \mempool{}'s implementation and discusses its performance trade-offs.


\section{Background \& Related Work}
\label{sec:relwork}

Today, multi-\gls{pe} \glspl{soc} are ubiquitous and cover a large spectrum in the trade-off among flexibility, performance, energy efficiency, and programmability.
In the following, we focus on the foundations, evolution, and state-of-the-art of the three main classes of scalable, multi-\gls{pe} architectures: systolic arrays, \glspl{cgra}, and shared-L1-memory manycores.
\rebuttal{A fair quantitative comparison among such architectures is challenging due to differences in technology, core count, and evaluated workloads. Nevertheless, in the following, we provide reference numbers for a high-level comparison.}
A qualitative comparison is reported in \Cref{tab:related_work}.

\subsection{Systolic Array Architectures}
\label{subsec:rwork_systolic}

Systolic array architectures were proposed by Kung in 1982 in the field of high-performance acceleration~\cite{Kung82}.
In a systolic architecture, neighboring \glspl{pe} directly transfer data to linked \glspl{pe} across a fixed-topology network, executing the processing in lockstep. Memory accesses only occur at the boundaries of the \gls{pe} array.
A \gls{pe} interconnect topology matching the algorithm's dataflow greatly boosts the application's performance and energy efficiency. However, this ties each systolic array to the application it has been designed for.
As a matter of fact, systolic arrays usually target \gls{asic}~\cite{fornt2023energy} or domain-specific accelerator~\cite{Mei2003} implementations.
One prominent example is Google's \gls{tpu}~\cite{jouppi2017datacenter}, an accelerator used for deep learning and featuring \num{65536} 8-bit \gls{mac} units connected in a 2D-mesh systolic array. \rebuttal{The \gls{tpu} claims to reach a \SI{92}{\tera\ops} peak throughput for 8-bit \glspl{mac}, with an energy efficiency of \SI{2}{\tera\ops\per\watt} at \SI{700}{\mega\hertz}, in a \SI{28}{\nano\meter} technology.}

In addition to \emph{pure-systolic} designs, Kung introduced \emph{semi-systolic} architectures~\cite{Kung82} that allow for global communication to and from any \gls{pe} in addition to peer-to-peer, systolic communication.
A notable implementation of such a hybrid execution model is provided by Eyeriss~\cite{chen2016eyeriss}, a systolic accelerator for deep \glspl{cnn}. It features \num{168} \glspl{pe} that can communicate in a peer-to-peer fashion or collectively, through a \gls{noc}-based global buffer. Despite its communication flexibility, domain-specific specializations at the \gls{pe}, \gls{noc}, and memory level tie it to \glspl{cnn} workloads. \rebuttal{With these, it achieves \SI{100}{\percent} \gls{pe} utilization, with a throughput of \SI{33.6}{\giga\ops} at \SI{200}{\mega\hertz}.}

To the best of our knowledge, no massively parallel, general-purpose architectures supporting a hybrid execution model exist in the literature. As a matter of fact, implementing a global, all-to-all interconnect for multicast and gather communication is prohibitive in terms of scalability and cost. Efficient, hardwired solutions make the systolic topology even more application-specific, making it hard to justify their extra cost.

\colorlet{coloryes}{OliveGreen}
\colorlet{colorjein}{YellowOrange}
\colorlet{colorno}{OrangeRed}

\colorlet{colorverygood}{OliveGreen}
\colorlet{colorgood}{OliveGreen}
\colorlet{colorneutral}{Gray}
\colorlet{colorbad}{OrangeRed}
\colorlet{colorverybad}{OrangeRed}

\newcommand{\verygood}[1]{\textcolor{colorverygood}{#1}}
\newcommand{\good}[1]{\textcolor{colorgood}{#1}}
\newcommand{\neutral}[1]{\textcolor{colorneutral}{#1}}
\newcommand{\bad}[1]{\textcolor{colorbad}{#1}}
\newcommand{\verybad}[1]{\textcolor{colorverybad}{#1}}

\newcommand\circledsym[2]{%
  \adjustbox{height=1.15em,margin*=0 -1.25 -2.5 0}{%
    \tikz\node[circle,color=white,fill=#1,inner sep=.2pt,font=\bfseries]{#2};%
  }
}

\newcommand{\yes}{\circledsym{coloryes}{$\pmb\checkmark$}}
\newcommand{\no}{\circledsym{colorno}{$\pmb\times$}}
\newcommand{\jein}{\circledsym{colorjein}{$\pmb\approx$}}
\newcommand{\unknown}{\circledsym{colorneutral}{?}}

\newcolumntype{Y}{>{\centering\arraybackslash}X}
\newcolumntype{Z}{>{\raggedleft\arraybackslash}X}
\newcolumntype{R}{%
  >{\adjustbox{right=6.8em,angle=310,lap=-\width+1.5em}\bgroup}%
  c%
  <{\egroup}%
}
\newcolumntype{M}{%
  >{\adjustbox{right=6.8em,angle=310,lap=-\width+2.5em}\bgroup}%
  c%
  <{\egroup}%
}
\newcommand*\rot{\multicolumn{1}{R}}%
\newcommand*\rotm{\multicolumn{1}{M}}%
\newcommand{\tblrottitle}[1]{\rot{{#1}}}
\newcommand{\tblrottitlem}[1]{\rotm{{\makecell[cr]{#1}}}}

\newcommand{\underlinecenter}[2]{%
\setul{3pt}{.4pt}
\ul{\mbox{\hspace{#1}}#2\mbox{\hspace{#1}}}}

\begin{table}[tb]
  \caption{Qualitative comparison of our hybrid architecture with related, highly parallel designs.}
  \label{tab:related_work}
  \vspace{-0.4cm} 
  \setlength{\tabcolsep}{1pt}%
  \sisetup{range-phrase=--}%
  \center%
  \begin{tabularx}{\columnwidth}{@{}lYYYYYYY@{}}
    \toprule
        \parbox[t][5em][b]{1em}{Architecture}
      & \tblrottitle{Total \glspl{pe}}
      & \tblrottitlem{Programmable, \\ independent \glspl{pe}}
      & \tblrottitlem{General-purpose}
      & \tblrottitlem{Efficient P2P \\ communication}
      & \tblrottitlem{Dynamic \\ reconfiguration}
      & \tblrottitlem{Efficient \\ multicast/gather}
      & \tblrottitle{Open source}
    \\ %
    \midrule  
        Google TPU~\cite{jouppi2017datacenter}
      & 65536
      & \no{}
      & \no{}
      & \yes{}
      & \no{}
      & \no{}
      & \no{}
    \\ %
        Eyeriss~\cite{chen2016eyeriss}
      & 168
      & \yes{}
      & \no{}
      & \yes{}
      & \yes{}
      & \yes{}
      & \no{}
    \\ %
        HyCUBE~\cite{Karunaratne2017}
      & 16
      & \no{}
      & \no{}
      & \yes{}
      & \yes{}
      & \yes{}
      & \yes{}
    \\ %
        Pixel Visual Core~\cite{Redgrave2018}
      & 2048
      & \yes{}
      & \no{}
      & \yes{}
      & \jein{}*
      & \no{}
      & \no{}
    \\ %
        KiloCore~\cite{bohnenstiehl2017kilocore}
      & 1000
      & \yes{}
      & \yes{}
      & \no{}
      & \no{}
      & \no{}
      & \no{}
    \\ %
        Celerity~\cite{davidson2018celerity}
      & 496
      & \yes{}
      & \yes{}
      & \no{}
      & \no{}
      & \no{}
      & \yes{}
    \\ %
        Epiphany-V~\cite{olofsson2016epiphany}
      & 1024
      & \yes{}
      & \yes{}
      & \no{}
      & \no{}
      & \no{}
      & \no{}
    \\ %
        RC64~\cite{Ginosar2016}
      & 64
      & \yes{}
      & \yes{}
      & \no{}
      & \no{}
      & \no{}
      & \no{}
    \\ %
        Gap9~\cite{GreenWavesTechnologies2021}
      & 9
      & \yes{}
      & \yes{}
      & \no{}
      & \no{}
      & \no{}
      & \jein{}\dag
    \\ %
        \mempool{}~\cite{riedel2023mempool}
      & 256
      & \yes{}
      & \yes{}
      & \no{}
      & \no{}
      & \yes{}
      & \yes{}
    \\ %
        \textbf{This work}
      & 256
      & \yes{}
      & \yes{}
      & \yes{}
      & \yes{}
      & \yes{}
      & \yes{}
    \\ %
    \bottomrule
  \end{tabularx}
  \scriptsize
  \raggedright
  \\[1mm]
  * Only coarse-grain, through the \gls{noc}, and limited by the ring-shaped topology.\\
  \dag\ Closed source based on open source.
\end{table}

\subsection{Coarse-grained Reconfigurable Architectures}
\label{subsec:rwork_cgra}

A step towards higher flexibility is represented by \acrfullpl{cgra}, which feature \glspl{pe} and interconnect networks with a higher degree of configurability~\cite{Podobas2020}.
HyCUBE, for example, implements a \gls{cgra} with a 2D mesh topology~\cite{Karunaratne2017} that can be configured to connect, at each cycle, any two \glspl{pe} with single-cycle access latency. \rebuttal{The 16-\gls{pe}, \SI{28}{\nano\meter} HyCUBE implementation achieves \SI{63}{\mega\ips\per\watt} at \SI{704}{\mega\hertz}.} Its scalability, however, is confined to a few tens of \glspl{pe} due to the combinational bypasses employed to implement the reconfigurable 2D mesh.
Google's Pixel Visual Core~\cite{Redgrave2018} features eight \num{256}-\gls{pe} stencil processors communicating in a ring interconnect through a \gls{noc}.
This toroidal configuration is highly efficient with neighborhood-based pixel computation, \rebuttal{for which it boasts a peak throughput of \SI{1}{\tera\ops} at \SI{426}{\mega\hertz}, with a power consumption of \SI{4.5}{\watt} when implemented in a \SI{28}{\nano\meter} technology.} The topology configurability, however, is limited to the \gls{noc}-based ring, poorly mapping general-purpose workloads.

Compared to an application-specific systolic array, a \gls{cgra}-based architecture trades \gls{pe} and interconnect energy efficiency for higher flexibility.
However, the possible topologies are still dictated by the \gls{cgra}'s fabric, limiting the communication patterns that can be efficiently implemented among the \glspl{pe}. \Gls{pe} functionalities, although broader, are usually limited to a specific application class, and the lack of a unified programming language limits programmability.
In an attempt to bridge this gap, Farahani \etal embed in the MIPS32 datapath a 4-\gls{pe} \gls{cgra} which automatically reconfigures based on the instructions detected at runtime~\cite{Farahani2020}. However,
this approach requires a high hardware overhead and is exclusively conceived for serial computation.
\rebuttal{As exemplified in \Cref{fig:pareto}, our hybrid architecture does not target the performance and energy efficiency of state-of-the-art, domain-specific systolic arrays and \glspl{cgra}, which can be up to one order of magnitude higher. It rather strives to accelerate a variety of regular workloads with high arithmetic intensity on general-purpose, manycore systems, striking a new trade-off point between flexibility and efficiency.}

\subsection{Manycore \& Shared-L1-Memory Clusters}
\label{subsec:rwork_shmem}

Manycore systems are general-purpose architectures integrating hundreds of cores, usually based on cache hierarchies, \glspl{noc}, and \glspl{spm}.
Rather than specialized functional units, manycore systems' \glspl{pe} are full-fledged programmable cores. Their communication happens via explicit memory accesses at user-specified addresses, which results in maximum flexibility.

Examples of state-of-the-art, scaled-up manycore systems are KiloCore~\cite{bohnenstiehl2017kilocore} and Celerity~\cite{davidson2018celerity}, featuring respectively 1000 and 496 logically connected cores. To tackle the massive scale-up, they exploit neighborhood-based interconnects, which imply up to 60 cycles for message routing across \glspl{pe}. \rebuttal{In a \SI{32}{\nano\meter} technology, at \SI{1.78}{\giga\hertz}, and \SI{13}{\watt}, KiloCore's achieves a \SI{138}{\giga\ops\per\watt} energy efficiency.}
Epiphany-V~\cite{olofsson2016epiphany} is another example of a manycore system scaling up to 1024 cores, organized in a \gls{noc}-based 2D mesh.
\rebuttal{It targets an energy efficiency of \SI{75}{\giga\flops\per\watt}.}
Nevertheless, the \gls{noc} implies inter-core communication latencies proportional to the cores' relative position and up to tens of cycles. 
Therefore, acceptable performance figures require spatially-aware distribution of workloads and data, ultimately affecting the programmability and the flexibility of such distributed systems~\cite{Pathania2018}.

A highly flexible architectural pattern with a balanced efficiency trade-off that still achieves acceptable efficiency is the shared-L1 cluster. The most successful examples of such architectures are the \gls{sm} units in GPUs, like the NVIDIA Hopper's \glspl{sm} featuring 128 \glspl{pe}~\cite{NVIDIACorporation2022} \rebuttal{and boasting up to \SI{134}{\tera\flops} in terms of 32-bit operations at \SI{1.8}{\giga\hertz}. Such a performance, however, requires a power consumption up to \SI{400}{\watt} and sacrifices instructions stream diversity.} %
One key challenge for shared-L1 clusters is scaling toward hundreds of cores and more. Most shared-L1 clusters are in the range of the tens of cores, like the RC64~\cite{Ginosar2016}, or GAP9~\cite{GreenWavesTechnologies2021}, but \mempool{}~\cite{riedel2023mempool} scaled the core count up to 256.
\mempool{} is a \riscv{}-based general-purpose manycore system whose independently programmable 256 cores share a large pool of L1 \gls{spm} through a low-latency, hierarchical, crossbar-based interconnect. 
Without contentions, every core can access any memory address in at most five cycles, which the cores can hide thanks to a non-blocking \gls{lsu} and a scoreboard for tracking dependencies~\cite{zaruba2020snitch}. This removes the need for spatial mapping of algorithms and data, greatly boosting programmability.

However, \riscv{}'s load-store architecture implies a runtime overhead for the explicit communication through memory. For regular kernels, such as matrix multiplications or convolutions, \mempool{} spends up to \SI{35}{\percent} of the runtime on non-compute instructions and synchronization~\cite{riedel2023mempool}. Nevertheless, their extremely regular dataflow can be exploited to optimize data and movement, similarly to systolic architectures~\cite{Kung82}. Implementing systolic dataflow in the L1 memory of a \mbox{\riscv{}-based} system is usually highly inefficient because of explicit loads and stores and synchronization overhead on software FIFOs.

Our hybrid systolic-shared-memory approach addresses these shortcomings: while retaining the same flexibility of a shared-memory architecture, it achieves significantly higher efficiency when workloads can be executed in a systolic fashion. The proposed hardware extensions allow the data to seamlessly move between cores through the memory-mapped queues, parallel to the computation, and in an energy-efficient way. This overcomes the limitations of the load-store architecture and frees the cores of explicit memory and synchronization operations for systolic kernels.
\rebuttal{Our architecture achieves up to \SI{208}{\giga\ops\per\watt} for 32-bit operations at \SI{600}{\mega\hertz}, with a throughput of \SI{227}{\giga\ops} and a power consumption of about \SI{1.1}{\watt}, qualitatively positioning itself as in \Cref{fig:pareto}.}


\section{Hybrid Architecture}
\label{sec:architecture}

This section presents our hybrid systolic-shared-memory architecture.
The contribution presented hereby is twofold: Firstly, we motivate and discuss the application of a systolic overlay to a shared-memory cluster. 
Secondly, we discuss the two low-overhead \gls{isa} extensions that enable high performance and energy efficiency for the hybrid architecture.

In the following, we use the words \emph{core} and \emph{\gls{pe}} alternately. While they both refer to the same hardware block, we employ \emph{core} when in the context of a shared-memory cluster and \emph{\gls{pe}} to specifically connote the processing unit of a systolic array.

\subsection{Hybrid Architectural View}
\label{subsec:hybrid_arch}

\begin{figure}[tb]
    \centering
    \includegraphics{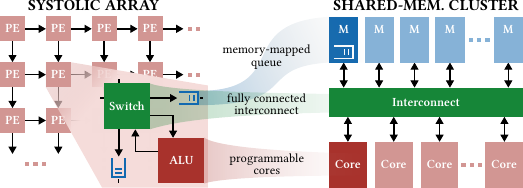}
    \caption{Example mapping of a purely systolic \by{$N$}{$N$} 2D-mesh architecture (\emph{left}) over a generic a $N^{2}$-core L1-shared cluster (\emph{right}): the cluster's $N^{2}$ cores can be viewed as a \by{$N$}{$N$} array of \glspl{pe} connected to their neighbors through memory-mapped queues.}%
    \label{fig:architecture_overview}
\end{figure}

From a high-level perspective, a shared-L1-memory cluster comprises several cores and a pool of banked shared memory. As any core can access any memory address, the memory acts as a de-facto all-to-all communication channel.
By mapping a user-defined number of FIFO queues to the shared memory, synchronous peer-to-peer links among any two cores can be implemented.
If all cores have low-latency access to any memory address, then the memory-mapped queues can be exploited to efficiently emulate any systolic topology among the \glspl{pe}, up to the shared memory physical limit, implementing \emph{systolic links}, as in \Cref{fig:architecture_overview}.
If the memory-mapped queues are software-configurable, the systolic topology can be dynamically reconfigured at runtime.

\rebuttal{
However, the key advantage of purely systolic arrays is that inter-\gls{pe} communication is implicit by the hardwired systolic links.
In contrast, systolic dataflow via memory-mapped queues seemingly intensifies the L1-memory access, decreasing the cluster's utilization and energy efficiency.
In truth, in our hybrid architecture, the systolic and the shared-memory models mutually benefit each other. As a matter of fact, while support for a systolic dataflow allows the shared-memory cluster to map regular workloads more efficiently, the cluster's properties reciprocally boost the systolic execution:
}

\begin{itemize}
    \item Shared-L1-memory manycore clusters generally resort to a banked \gls{tcdm}. Equally distributing the memory-mapped queues across all banks optimizes data locality and memory access contentions, lowering access latencies.
    \item The \glspl{pe} of our hybrid architecture are fully programmable cores provided with a register file, which allows them to maximize data reuse by tiling the computation.
    This effectively reduces the memory accesses to exchange intermediate systolic results.
    \item As the systolic topology is fixed before the computation phase, the memory addresses of the queues are constant. Therefore, a systolic execution model relieves the cores of load and store addresses computation, increasing functional unit utilization.
    \item Scaled-up manycore clusters employ a \gls{numa} architecture to make the all-to-all interconnect feasible.
    While we assume a low-latency access to the L1, minimizing systolic links across remote core further improves the hybrid architecture's efficiency.
    \item \rebuttal{Concurrently feeding (or retrieving) data to (or from) multiple \glspl{pe} is a great benefit to systolic execution~\cite{Kung82}. The shared memory intrinsically implements an efficient mechanism of multicast and gather communication through common, concurrent load and store operations, allowing hybrid systolic-shared-memory execution models.}
\end{itemize}
\rebuttal{
We employ these devices in \Cref{sec:kernels}, exploring the enabled hybrid execution models. Their actual effectiveness in improving the cluster's utilization is demonstrated in \Cref{sec:evaluation}.
}

It is noteworthy that, as memory-mapped queues can be implemented without affecting any other feature of the cluster, our hybrid architecture accelerates regular, compute-intensive workloads without any regression on other workload types.

\subsection{Software Systolic Emulation}
\label{subsec:systolic_emulation}

\begin{figure*}[thb]
    \centering
    \includegraphics{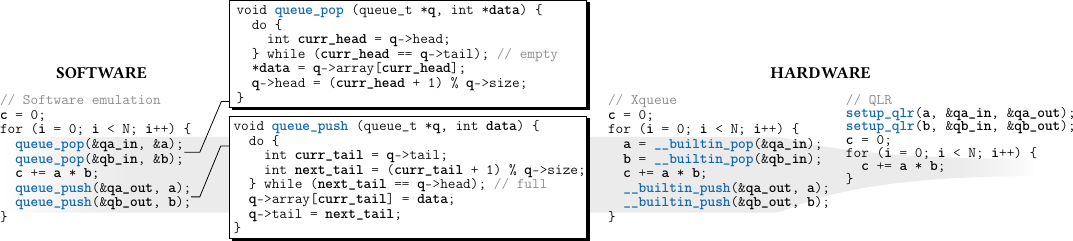}
    \caption{Simplified code executed on the hybrid architecture's \glspl{pe} to implement a matrix multiplication. With the systolic \emph{Software emulation} (left), the \gls{pe} accesses the queues defined by the four \texttt{queue\_t} structures through function calls encompassing explicit queue bookkeeping and access. The \emph{\xqueue{}} hardware extension (middle) replaces such function calls with single instructions only requiring the queues' address.
    With the further \emph{\gls{qlr}} extension (right), after a minimal set-up phase, communication is performed implicitly and queue instructions are totally elided.
    }
    \label{fig:code}
\end{figure*}

To implement our hybrid architectural view, we develop a software runtime to emulate systolic topologies in an L1-shared cluster.
The systolic runtime is a software library aiding memory-mapped queue allocation and management. The programmer controls the queues' size, number, and location.

The runtime implements a push and a pop function for memory-mapped queue access.
As in \Cref{fig:code}, \qpush{} and \qpop{} implement the queues as circular buffers. The functions explicitly handle head and tail bookkeeping, actively waiting on the pointers until the FIFO boundary conditions are met.
This synchronizes all cores without any explicit barrier, implementing the pulsed dataflow of systolic arrays.
%
To establish a systolic topology, it is enough to allocate the desired queues in the shared memory and assign them to producer and consumer cores.
Assigning a single consumer and producer couple to each queue guarantees the absence of race conditions on the queue pointers.

While the systolic runtime can be used to implement a systolic communication network among the cores, concurrent memory access to generic addresses can still be employed for multicast and gather global communication. The usage of \qpush{} and \qpop{} can indeed be alternated with normal load and store instructions to obtain hybrid systolic-shared-memory execution models, boosting systolic algorithms performance beyond purely systolic approaches.

With the proposed runtime, systolic topologies can be implemented on any shared-memory multi-core architecture without additional hardware.
Unfortunately, software-managed queues require tens of instructions for each push and pop operation, including multiple additional memory accesses for queue bookkeeping. While being a useful exploration tool, a software-emulated systolic array incurs extensive control overhead, hence poor performance.
In the next section, we discuss two hardware extensions tackling this issue.

\subsection{Hardware Extensions}
\label{subsec:hardware_extensions}

In a traditional systolic execution, data automatically flows through \glspl{pe}. On the other hand, a shared-L1-memory cluster requires explicit memory accesses. We propose two lightweight hardware extensions to fill this gap, namely \xqueue{} and \Acrfullpl{qlr}. This section presents their specification. Their reference, open-source implementation is discussed in \Cref{sec:implementation}.


\subsubsection{\xqueue{}}
\label{subsubsec:xqueue}

The \xqueue{} extension overcomes the large overhead of manual queue bookkeeping by enhancing the \riscv{} \gls{isa} with two memory instructions, namely, \texttt{q.push} and \texttt{q.pop}.
In one single instruction, \xqueue{} hardware takes care of queue boundary checks, head/tail pointer updates, and queue access.
\Cref{fig:code} illustrates the benefits of \xqueue{} in terms of computation/control ratio.

\xqueue{} stores head and tail pointers as internal registers so that their updates do not generate memory accesses.
Based on those pointers, \xqueue{} can stall push and pop operations that cannot be executed immediately. By integrating \xqueue{} with the cores' scoreboard, the systolic links can exert backpressure on the cores in case of a failed queue boundary check,
which implements implicit synchronization among \glspl{pe}.

\xqueue{} supports a user-defined number of configurable-size queues, mapped at any address of the shared memory.
However, we can greatly simplify the queue manager by parametrizing the number, the size, and the location of the queues.
This results in negligible hardware overhead, nevertheless allowing a configurable degree of flexibility for the systolic topologies.


\subsubsection{Queue-linked Registers}
\label{subsubsec:qlr}

The \xqueue{} extension significantly reduces the number of control instructions required to manage inter-\gls{pe}, queue-based communication.
Building on top of \xqueue{}, the \gls{qlr} extension makes fast queue access also autonomous, eliding those instructions. The rightmost snippet in \Cref{fig:code} illustrates this concept.

Analogously to a \gls{udma} controller deeply embedded in the \gls{pe} microarchitecture~\cite{schuiki2020stream}, \glspl{qlr} take care of data transfers between the L1 memory and the \gls{pe}'s register file. \glspl{qlr}' purpose is to track accesses to pre-defined registers: depending on their configuration, they implicitly turn the detected register reads into queue pops and register writes into queue pushes, making inter-core communication autonomous. While doing this, they hide interconnect and memory latencies with internal FIFOs.
In order to do so, a \gls{qlr} requires 4 main interfaces:
\begin{itemize}
    \item \emph{Memory interface}: A \gls{qlr} can issue memory transactions to autonomously request \texttt{q.push} and \texttt{q.pop} operations.
    \item \emph{Register file write-back}: To operate transparently with respect to the \gls{pe} architecture, a \gls{qlr} writes back new data to the register file after every pop. This leaves the rest of the datapath unaffected.
    \item \emph{Instruction decoder snooping}: The \gls{qlr} snoops the register file's write and read ports to detect when its register is accessed and whether to request a \texttt{q.push} or a \texttt{q.pop}.
    \item \emph{Scoreboard interface}: If a \gls{pe}'s \gls{qlr} requires new data when its input queue is empty, a \gls{raw} hazard can occur. Similarly, a \gls{waw} can happen when a \gls{pe} issues a push to a full queue.    
    Access to the cores' scoreboard allows \glspl{qlr} to handle queue-related hazards by propagating the backpressure from the systolic links and accordingly generating stalls.
\end{itemize}

\noindent A parametrizable number of \glspl{qlr} can be tightly integrated with the cores, sharing their memory port and register file interface. Each \gls{qlr} is tied to one register in the register file.
A \gls{qlr} can be in three operating modes: 1) \emph{incoming}, 2) \emph{outgoing}, or 3) \emph{in-out}, depending on whether it pops operands, pushes operands, or pops and directly forwards operands to another queue. An incoming \gls{qlr} can be configured to provide the same operand multiple times before popping the next element from the queue, which automates data reuse.
Each core's \gls{qlr} can be programmed via private memory-mapped \glspl{csr} by setting the address of its related memory-mapped queue, its operating mode, and, in case of incoming \gls{qlr}, the degree of data reuse.

\glspl{qlr} allow the creation of a queue network that, once configured, runs entirely autonomously. However, the number of available \glspl{qlr} per core is a hardware parameter. This creates an upper bound only for the number of implicitly accessed queues. Should a \gls{pe} need additional systolic links, explicit \xqueue{} instructions can always be used.


\section{Hybrid Systolic-Shared-Memory \mempool{}}
\label{sec:implementation}

In this section, we describe the implementation of the hybrid architecture proposed in \Cref{sec:architecture}. In particular, we implement the \xqueue{} and the \gls{qlr} extensions in \mempool{}~\cite{riedel2023mempool}, the shared-L1-memory manycore cluster presented in \Cref{subsec:rwork_shmem}. We evaluate this implementation in \Cref{sec:evaluation}.

\subsection{Hybrid Architectural View}
\label{subsec:mempool_hybrid_arch}

\mempool{} features 256 \riscv{} cores hierarchically interconnected to \SI{1}{\mega\byte} of shared L1 \gls{spm}. The shared memory is partitioned into 1024 banks~\cite{riedel2023mempool}.
\mempool{}'s \gls{pe} is the \mempool{} \gls{cc}, composed of a 32-bit \riscv{} Snitch processor coupled with a \gls{dsp} accelerator~\cite{zaruba2020snitch}. Four \glspl{cc}, 16 banks of \gls{spm}, and a shared instruction cache form a \mempool{} tile. 16 tiles form a \mempool{} group, and four groups form the full \mempool{} cluster. In the absence of contentions, the cores can access memory banks in the same tile (i.e., \textit{local banks}) in one cycle and remote banks within five cycles.

According to our hybrid architectural view, any of the 256 cores can communicate through systolic links via the shared L1 memory. However, due to \mempool{}'s hierarchical interconnect, accesses to local memory banks are faster and more energy efficient than the ones directed to remote tiles or groups. Therefore, as illustrated in \Cref{sec:kernels}, minimizing systolic links based on remote memory transactions plays an important role in the design of systolic dataflows for \mempool{}.

\subsection{\xqueue{}}
\label{subsec:mempool_xqueue}

\begin{figure}[t]
    \centering
    \includegraphics{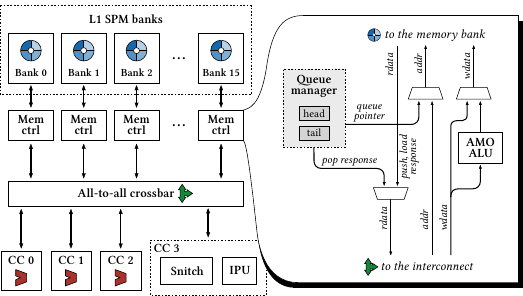}
    \caption{Simplified overview of the \mempool{} tile, on the left, and of the \xqueue{}-extended memory controller, on the right. In the hybrid view of \mempool{}, each \gls{cc} is a \gls{pe} of the systolic array, and each \gls{spm} bank reserves space for a memory-mapped queue handled by its \textit{queue manager}.}
    \label{fig:xqueue}
\end{figure}

In \mempool{}, each of the 1024 \gls{spm} banks features a small memory controller. To implement \xqueue{}, we extend it with minimal hardware to natively support memory-mapped queue management. \texttt{q.push} and \texttt{q.pop} have a similar signature as \riscv{} \glspl{amo}, already supported in \mempool{}. Therefore, we leverage \glspl{amo}' control path to integrate \xqueue{} with negligible overhead.
\Cref{fig:xqueue} outlines the structure of \mempool{}'s \xqueue-extended memory controller.

In particular, we enhance the memory controller with a head and tail register for the boundary conditions check.
On a queue transaction, if the FIFO boundary conditions are satisfied, the queue manager translates the request into a read or write to the correct address based on the queue pointers. When popping an empty queue, the response is withheld until another core pushes a new element. During this time, the memory controller stays fully functional for other non-queue-related transactions.
On the other hand, when pushing to a full queue, the operand is buffered to make the push non-blocking.
Similarly to the software implementation in \Cref{fig:code}, \xqueue{} implements memory-mapped queues as circular buffers: this simplifies the hardware for the boundary checks at the cost of a wasted queue entry. Nevertheless, we employ the empty slot to buffer the operand of stalled push operations. A pop freeing a spot triggers the outstanding push response.
For minimal hardware overhead, \xqueue{} supports only one outstanding push or pop, and backpressure is exerted on the issuing cores at the occurrence of further queue requests.

For the evaluation of \Cref{sec:evaluation}, we configure \mempool{} to have one queue manager per memory bank. This makes 1024 hardware-accelerated queues available to implement systolic links. For each queue, we allocate four 32-bit entries.
As in \Cref{subsec:hybrid_arch}, having one queue per bank minimizes the memory contention due to systolic communication since only one producer and one consumer \glspl{pe} access each bank.

\subsection{Queue-linked Registers}
\label{subsec:mempool_qlr}

\begin{figure}[t]
    \centering
    \includegraphics{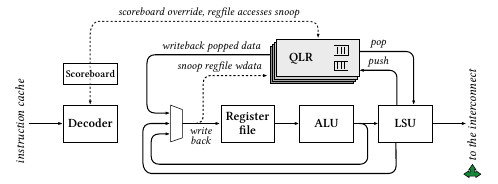}
    \caption{Schematic of the Snitch core extended with four \glspl{qlr}. \glspl{qlr} tap into the register file ports and the instruction decoding logic to detect register accesses and intercept newly written data. They also have access to the scoreboard, to the register file write-back path, and to the \gls{lsu}.}%
    \label{fig:qlr}
\end{figure}

\xqueue{} provides all 256 cores with fast access to 1024 queues mapped in \mempool{}'s \gls{spm}. Complying with \mempool{}'s banking factor of four, we instantiate four \glspl{qlr} per core. Each \gls{qlr} allows a core to implement an autonomous systolic link that leverages \xqueue{}'s queue manager.

We integrate support to \glspl{qlr} directly in the Snitch cores to facilitate their interface to the register file.
\mempool{}'s \glspl{qlr} are tied to the temporary registers from \texttt{t0} to \texttt{t3} of the \riscv{} calling convention.
\glspl{qlr} require the ability to issue \texttt{q.push} and \texttt{q.pop} memory requests: to avoid an additional, costly memory interface, we route \glspl{qlr} requests through Snitch's \gls{lsu}.
\Cref{fig:qlr} shows the architecture of the \gls{qlr} extension implementation.


\section{Hybrid Execution Model}
\label{sec:kernels}

Bridging the systolic and the shared-memory execution models opens unprecedented software design trade-offs supported by the hybrid architecture.
In this section, we leverage \mempool{} to illustrate the \gls{dse} for several \gls{dsp} kernels with diverse arithmetic intensities, namely matrix multiplication, 2D convolution, complex \gls{fft}, and dot product.

Note that the specific implementation of \mempool{}'s systolic links through software emulation, \xqueue{}, or \glspl{qlr} is orthogonal to the software execution model. We, therefore, postpone its discussion to \Cref{sec:evaluation}.

\begin{figure*}[th]
    \centering
    \includegraphics{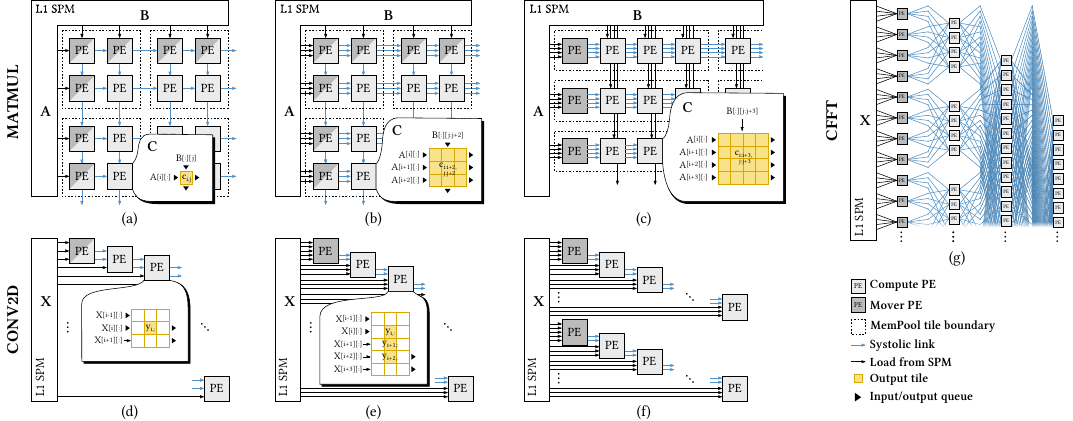}
    \caption{\rebuttal{The hybrid systolic-shared-memory topologies for the \emph{matmul}, \emph{conv2d}, and \emph{cfft} benchmarks implemented for \mempool{}. For \emph{matmul} and \emph{conv2d}, only the main incremental versions are depicted. For \emph{cfft}, our radix-4 Cooley-Tukey implementation is shown. All schemes are described in detail in \Cref{sec:kernels}.}}%
    \label{fig:systolic_topologies}
\end{figure*}

\rebuttal{

\subsection{Hybrid \acrlong{dse}}
\label{subsec:kernels-dse}

In the following, we collect the main insights in the exploration of the hybrid design space, revolving around two aspects: dataflow and core-\acrshort{pe} mapping.

\subsubsection{Dataflow}
\label{subsubsec:dse-dataflow}
The dataflow specifies the way that inputs, partial results, and outputs move across the \glspl{pe}.
With a purely systolic dataflow, only \glspl{pe} at the array boundary access memory for input/output.
On the other hand, the shared-memory model empowers each \gls{pe} with independent, concurrent access to memory through load/store instructions.
In particular, concurrent load operations can implement broadcast communication to increase \glspl{pe}' input capabilities beyond the limits of the available systolic links, allowing larger data reuse. Additionally, concurrent stores can efficiently perform global gather operations for the final output phase, which avoids lengthy shifts to the boundary \glspl{pe} when an output-stationary systolic dataflow is employed~\cite{Kung82}.
However, such a hybrid dataflow requires explicit load/store instructions during systolic execution, which reduces functional unit utilization and might be subject to memory contentions. Therefore, a trade-off exists.

\subsubsection{Core-\acrshort{pe} mapping}

From the perspective of the individual \glspl{pe}, the shared-memory manycore cluster features fully programmable cores with register files.
Therefore, their resources can be better exploited to increase data reuse and reduce the amount of required communication among \glspl{pe}. This is achieved by coalescing batches of computation over each core rather than individual operations.
For each hot-loop iteration, this approach requires a higher number of input operands that can be obtained with the abovementioned hybrid dataflow. However, data reuse can be increased only up to the limit dictated by the register file size to prevent detrimental stack spills.

From the perspective of the full array, the systolic links connect multiple \glspl{pe} in directed chains. Depending on the adopted systolic topology, a \gls{pe} might be part of multiple systolic chains, as is the case of a 2D-mesh topology with horizontal and vertical chains.
In systolic chains, the slowest \gls{pe} dictates the speed of all other \glspl{pe}.
As a matter of fact, if the downstream \glspl{pe} cannot consume the incoming operands fast enough, they exert backpressure on the upstream \glspl{pe}. The same happens if upstream \glspl{pe} cannot produce data at the pace of downstream \glspl{pe}, which get starved.
It is, therefore, crucial to balance the \glspl{pe}' workloads~\cite{sun2023sense}.
In particular, in addition to the computation scheduled on inner \glspl{pe}, boundary \glspl{pe} also map memory accesses for the systolic array input/output.
This might create bottlenecks.
To restore maximum throughput, it is usually enough to relieve the boundary \glspl{pe} of computation, in fact, obtaining \textit{mover \glspl{pe}} and \textit{compute \glspl{pe}}.
However, removing \glspl{pe} from computational tasks decreases the maximum achievable throughput of the cluster. 

As in a pipeline, the length of the systolic chains also influences the systolic array's performance~\cite{yuzuguler2023scale}. Additionally, given the number of cores in the cluster, the chain length is inversely proportional to the number of chains, with every chain requiring its head core to be a mover \gls{pe}.
Long systolic chains create deeper dependencies among \glspl{pe}, with higher utilization penalty in case of stall propagation across the programmable cores, and longer transient times before the steady-state throughput is reached. On the other hand, while shorter chains address these issues, they imply a higher number of mover \glspl{pe}, which decreases the cluster's maximum throughput.

Finally, the relative position of the \glspl{pe} connected by systolic links is also a parameter to optimize. As mentioned in \Cref{subsec:hybrid_arch}, our hybrid architecture assumes low-latency access to any shared-memory address, which removes the need for algorithm spatial mapping. However, a certain degree of optimization is still possible due to the \gls{numa} architecture. By minimizing the number of \textit{remote systolic links} (i.e., the links crossing \mempool{} tile or group boundaries), the latency and energy efficiency of \glspl{pe} communication can be optimized.
}

\subsection{Matrix Multiplication}
\label{subsec:matmul}

With \matmul{}, we denote a matrix-matrix multiplication kernel that computes $C = AB$.
In \matmul{}, many \glspl{pe} require the same rows and columns from, respectively, $A$ and $B$. Therefore, we adopt an output-stationary dataflow, where input operands from $A$ and $B$ are pushed through the systolic array while partial sums are accumulated within each \gls{pe}. At the end of each systolic iteration, all \glspl{pe} concurrently store their assigned $C$ tile back to memory.

The most intuitive mapping of \matmul{} over \mempool{}'s 256 cores, as in \Cref{fig:systolic_topologies}(a), considers the cluster as a \by{16}{16} grid of \glspl{pe}, with each \gls{pe} assigned with one single element of $C$ to compute.
Since the number of vertical and horizontal systolic links is balanced, we map \mempool{} tiles as \by{2}{2} grids of \glspl{pe}, which minimizes the number of remote systolic links.
%
%
To improve utilization, each \gls{pe} can be programmed to process \by{2}{2} tiles of $C$ instead of a single \gls{mac}, fully exploiting all four available memory-mapped queues per core.
%
%
For an even higher degree of data reuse, we can unbalance the vertical and horizontal systolic links: this allows the processing of \by{3}{3} tiles of $C$ on each \gls{pe}, as in \Cref{fig:systolic_topologies}(b). In this case, three links stream $A$'s rows horizontally, concurrently pushing all rows, while a fourth link streams $B$'s columns vertically, pushing each column's element sequentially. This approach can be pushed up to the point where the sequential stream of $B$ elements becomes the bottleneck.


Data reuse can be pushed even further with a hybrid dataflow where the elements of four rows from $A$ are streamed horizontally, while elements from $B$ columns are independently loaded from the shared memory by each \gls{pe}.
With this concept, each \gls{pe} can efficiently process \by{4}{4} tiles of $C$. 
As this hybrid dataflow only uses horizontal systolic links, the topology minimizing remote links maps  \mempool{} tiles as \by{1}{4} grids of \glspl{pe}.

In the topologies explored so far, the 16 left-boundary \glspl{pe} are a bottleneck as they act both as mover and compute \glspl{pe}.
By relieving them of the \glspl{mac} computation, we can rebalance the workload across the horizontal systolic chains, as in \Cref{fig:systolic_topologies}(c).
As each horizontal chain needs its first \gls{pe} to be a mover \gls{pe}, a trade-off unveils about the optimal number of chains.

\subsection{2D Convolution}
\label{subsec:conv}

With \emph{conv2d}, we denote the 2D convolution of an image, namely the $\by{M}{N}$ matrix $X$, with a \by{3}{3} kernel $K$, resulting in a zero-padded $\by{M}{N}$ matrix $Y$.
Regarding dataflow, we adopt a weight- and output-stationary convolution algorithm, with the rows from $X$ shifting horizontally through a chain-like systolic array topology. This allows optimal data reuse of $X$ and an efficient output phase leveraging concurrent store instructions, as in the \matmul{} case.
The first core of the chain accesses memory to load the required $X$ sub-column, forwarding it to the following \glspl{pe}. The inner \glspl{pe} exploit hybrid access to their assigned input rows.

\Cref{fig:systolic_topologies}(d) depicts an intuitive topology where all cores are connected by systolic links in a 256-\gls{pe} chain. Each \gls{pe} $i$ is assigned to compute the $i$-th output row, and, at each iteration, it is fed with a \by{3}{1} sub-column of $X$. Input rows $i-1$ and $i$ are popped from the previous \gls{pe}, while row $i+1$ is loaded from memory. The input rows required by the following \glspl{pe}, namely $i$ and $i+1$, are then forwarded through the outgoing systolic links.
The natural evolution of this topology is to balance the array by relieving the head \gls{pe} of computational tasks.

Data reuse can be improved by feeding each compute \gls{pe} with larger sub-columns of $X$. With \by{5}{1} input sub-columns, three $Y$ output rows per \gls{pe} can be computed. In this configuration, illustrated in \Cref{fig:systolic_topologies}(e), a \gls{pe} $i$ pops rows $i-1$ and $i$ from two incoming systolic links, while rows $i+1$, $i+2$, and  $i+3$ are loaded from memory. Rows $i+2$ and $i+3$ are also forwarded to the outgoing systolic links. This same principle can be applied up to the limit of maximum register file occupation.

To further increase performance, the long systolic chain of this topology can be divided into multiple independent sub-chains, each with its own head mover \gls{pe}, as in \Cref{fig:systolic_topologies}(f). This opens the hybrid \emph{conv2d} to a range of trade-offs analyzed in \Cref{sec:evaluation}.

\subsection{Complex FFT}
\label{subsec:cfft}

With \textit{cfft}, we denote a 256-point \gls{dft} processing the input array $X$ of 256 complex values.
To implement the \gls{fft}, we select a radix-4 \gls{dit_adj} Cooley–Tukey \gls{fft} approach~\cite{johnsson1992cooley}, which presents several advantages for our hybrid architecture.
In terms of core-\gls{pe} mapping, a radix-$r$ computation increases the granularity of the parallelized operation over each \gls{pe}, exploiting the full programmability of \mempool{}'s cores and improving data reuse. In particular, the radix-4 algorithm fully utilizes the four available systolic links per core.

The systolic topology of our Cooley–Tukey \gls{fft} is depicted in \Cref{fig:systolic_topologies}(g). With 256 complex values to process, the dataflow is divided into four stages, each including 64 \glspl{pe}. The stages are pipelined and implicitly synchronized by means of the systolic links. Therefore, at steady state, four independent \glspl{fft} are concurrently processed in the \mempool{} cluster.
We leverage a weight-stationary dataflow so that each of the four 64-\gls{pe} groups is statically tied to a stage. This makes the twiddle factors constant for each \gls{pe}, which allows their pre-loading. For the same reason, the destination \glspl{pe} addresses and shuffling order of each stage's output data are constant and can be computed in advance.

In a radix-4 \gls{dit_adj} \gls{fft} butterfly, the twiddle factors of the first stage are equal to one. This particularly suits our topology as it relieves the \glspl{pe} of the first stage from performing \gls{mac} operations, making it possible to schedule the input loads from L1 without creating a bottleneck.
The input array is accessed in a digit-reverse order so that the output samples can be directly stored in the correct, sequential order during the last stage.
The rest of the stages normally perform the recursive processing of their radix-4 butterfly, with each \gls{pe} receiving four inputs from the previous stage and forwarding four outputs to the next stage through the systolic links. The fourth stage additionally writes back the \gls{fft} output to the shared memory.

\rebuttal{
\subsection{Kernels with low arithmetic intensity}

Our hybrid systolic-shared-memory architecture accelerates regular, high-arithmetic-intensity kernels by better leveraging the data reuse opportunities they expose.
The systolic dataflow is beneficial when some data, either input operands or partial results, can be exchanged and reused among \glspl{pe}, reducing time- and energy-consuming explicit memory accesses while efficiently addressing synchronization. This is the case for \textit{matmul}, \textit{conv2d}, and \textit{cfft}.

Many relevant \gls{dsp} kernels have low arithmetic intensity, exposing little to no data reuse possibilities.
If an algorithm has no data reuse at all, no data can be exchanged through systolic links.
This is the case, for example, of \textit{axpy}, which requires two new, different operands for each single \gls{mac} operation, preventing any possible data sharing among \glspl{pe}.
Nevertheless, in such cases, the hybrid systolic-shared-memory architecture allows resorting to the purely shared-memory model without any performance degradation, as discussed in \Cref{sec:evaluation}.

In some other cases, very limited communication is still possible. When parallelizing a dot product kernel \textit{dotp} over \mempool{}'s 256 cores, two compute phases emerge. In the first phase, each \gls{pe} executes the dot product between its assigned segments of the two input arrays. Similarly to \textit{axpy}, this phase is completely memory-bound as two new loads are required for each \gls{mac}.
In the second phase, a reduction sum over all \glspl{pe}'s partial sums is performed. Despite the limited amount of communication among \glspl{pe}, this phase can be implemented through a radix-$r$ logarithmic reduction tree based on systolic links, with the improvements discussed in \Cref{sec:evaluation}.
}


\section{Evaluation}
\label{sec:evaluation}

\newcommand\mempoolbaseline{\mempool{}\ensuremath{_{\text{bl}}}}
\newcommand\mempoolxqueue{\mempool{}\ensuremath{_{\text{Xq}}}}
\newcommand\mempoolqlr{\mempool{}\ensuremath{_{\text{QLR}}}}
\newcommand\convbaseline{\benchmark{conv2d}{bl}{}}
\newcommand\convsystolicsw{\benchmark{conv2d}{sw}{}}
\newcommand\convsystolicxqueue{\benchmark{conv2d}{Xq}{}}
\newcommand\convsystolicqlr[1]{\ifx&#1&\benchmark{conv2d}{QLR}{}\else\benchmark{conv2d}{QLR}{#1}\fi}
\newcommand\matmulbaseline{\benchmark{matmul}{bl}{}}
\newcommand\matmulsystolicqlr[1]{\ifx&#1&\benchmark{matmul}{QLR}{}\else\benchmark{matmul}{QLR}{#1}\fi}
\newcommand\cfftbaseline{\benchmark{cfft}{bl}{}}
\newcommand\cfftsystolicqlr{\benchmark{cfft}{QLR}{}}
\newcommand\dotpbaseline{\benchmark{dotp}{bl}{}}
\newcommand\dotpxqueue{\benchmark{dotp}{Xq}{}}

In the previous sections, we discussed the dimensions of the hybrid systolic-shared-memory design space, namely the different implementations of a systolic link and the possible software execution models.
In the following, we evaluate such dimensions for the \mempool{} implementation.

\begin{figure}[t]
    \centering
    \includegraphics[width=\columnwidth]{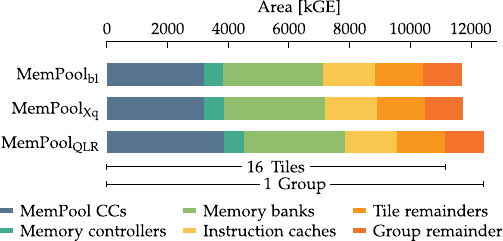}
    \caption{Area breakdown of the post-layout \mempool{} group, including 16 tiles. The tile and group remainder include the interconnect and the \gls{dma} area. The group has an area of \SI{11.7}{\mega\ge} for \mempoolbaseline{} and \SI{12.4}{\mega\ge} for \mempoolqlr{}.
    }
    \label{fig:group_area_links}
\end{figure}


\subsection{Methodology}
\rebuttal{\mempool{} is implemented in GlobalFoundries' 22FDX \gls{fdsoi} technology, targeting worst-case conditions (SS/\SI{0.72}{\volt}\kern-.15em/\SI{125}{\celsius})~\cite{riedel2023mempool}. We perform place-and-route with Synopsys Fusion Compiler 2022.03 and power simulations with Synopsys PrimeTime 2022.03. For power consumption estimations, we extract switching activities from post-layout gate-level simulations in typical conditions (TT/\SI{0.80}{\volt}\kern-.16em/\SI{25}{\celsius}) at \SI{600}{\mega\hertz}.}
%
As our hardware extensions have no impact outside of a tile, the following evaluation only considers one \mempool{} group, containing a total of 64 cores divided into 16 tiles.
With the \xqueue{} extension, a queue manager is implemented for each memory bank. For the \gls{qlr} extension, four \glspl{qlr} are integrated in each core.

\rebuttal{
For microarchitecture benchmarking, we employ four representative \gls{dsp} kernels, namely \textit{matmul}, \textit{conv2d}, \textit{cfft}, and \textit{dotp}. The \textit{matmul}, \textit{conv2d}, and \textit{dotp} benchmarks are implemented for 32-bit integer datatypes, while \textit{cfft} handles 32-bit complex numbers whose real and imaginary parts are 16-bit fixed-point values. The baseline benchmarks, described in Section 8.1 of \cite{riedel2023mempool} and Section V-A of \cite{bertuletti2023efficient}, are highly optimized for the purely shared-memory architecture. Their implementation is available in \cite{MemPoolRepo}. Their systolic variants evaluated in the following are presented in \Cref{sec:kernels}. All performance and power measurements come from cycle-accurate simulation.
}

For the following discussion, we carry out two complementary analyses as seen, for example, in \Cref{fig:stalls_links} and \Cref{fig:perfpower_links}.
The first analysis evaluates end-to-end benchmark executions with the same problem size, considering their relative performance in terms of functional resources utilization, \gls{ipc}, and stalls. \rebuttal{With a constant problem size, the relative utilization is directly proportional to the absolute runtime of the benchmark. To clearly highlight the impact of transient phases, only one kernel run is executed for each benchmark variant, without any double buffering.} The \emph{control} instructions, which negatively affect the \gls{pe} utilization, include load and store instructions, explicit queue accesses, and loop bookkeeping.
%
The second analysis considers the steady-state behavior of each benchmark, providing insights into peak performance, energy efficiency, and power consumption breakdown independently of the problem size.
For a fair comparison, each benchmark is executed on the minimal \mempool{} hardware that supports it.

\subsection{Evaluation of Systolic Link Implementations}
\label{subsec:eval_links}

In this section, we evaluate the different systolic link implementations discussed in \Cref{sec:architecture} and \Cref{sec:implementation}.
\Cref{fig:group_area_links} shows the area breakdown of the \mempool{} group in the three evaluated flavors, namely:
\begin{itemize}
    \item \mempoolbaseline{}: baseline \mempool{}, as in \cite{riedel2023mempool};
    \item \mempoolxqueue{}: \mempool{} extended with support for \xqueue{}, as in \Cref{subsec:mempool_xqueue};
    \item \mempoolqlr{}: \mempool{} extended with support for \xqueue{} and \glspl{qlr}, as in \Cref{subsec:mempool_qlr}.
\end{itemize}
The \xqueue{} and \qlr{} extensions have no impact outside of \mempool{} tiles. In particular, \xqueue{} only affects the memory controller of each \gls{spm} bank by implementing a queue manager, which determines an area overhead of about 0.5\% at the group level, or \SI{40}{\kilo\ge}. Each queue manager has an area of only \SI{150}{\ge}.
On the other hand, with \glspl{qlr}, each \mempool{} \gls{cc} gets about 20\% bigger, impacting the group with an additional \SI{700}{\kilo\ge} area. This sums up to an area increase of about 6\% with respect to \mempoolbaseline{}. Approximately \SI{2.5}{\kilo\ge} are required to enable a single register with \gls{qlr} capabilities.
\begin{figure}[t!]
    \centering
    \includegraphics{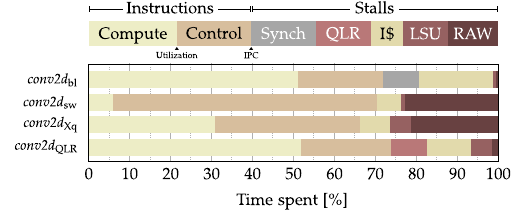}
    \vspace{-5.5mm}
    \caption{Relative performance results of the end-to-end \emph{conv2d} benchmark, based on the systolic links implementation.}
    \label{fig:stalls_links}
\end{figure}
\begin{figure}[t!]
    \centering
    \includegraphics[width=\columnwidth]{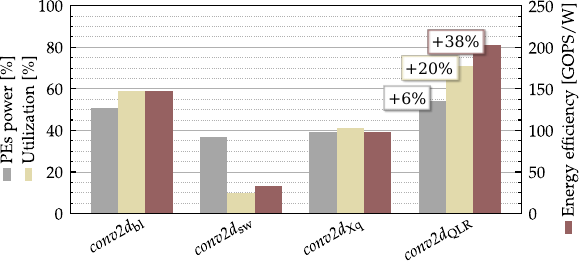}
    \vspace{-7mm}
    \caption{Steady-state analysis of the \emph{conv2d} benchmark, based on the systolic link implementation. The improvement with respect to \convbaseline{} is highlighted.}
    \label{fig:perfpower_links}
\end{figure}
\rebuttal{In terms of timing, all three versions achieve the same frequency of \SI{493}{\mega\hertz} in the worst-case corner, and can be clocked at \SI{600}{\mega\hertz} in the typical corner.} Therefore, the \xqueue{} and \gls{qlr} hardware does not cause any frequency degradation.

\Cref{fig:stalls_links} and \Cref{fig:perfpower_links} show the impact of the different systolic link implementations on the performance and energy efficiency of the \emph{conv2d} reference benchmark. For this analysis, we fix the systolic topology of the hybrid implementations to the one of \convsystolicqlr{7}, discussed in \Cref{subsec:eval_topology}.
We evaluate the following \emph{conv2d} flavors:
\begin{itemize}
    \item \convbaseline{}: the baseline \emph{conv2d} algorithm, highly optimized for the purely shared-memory architecture~\cite{riedel2023mempool,MemPoolRepo};
    \item \convsystolicsw{}: hybrid systolic-shared-memory \emph{conv2d}, implementing systolic links through software-emulated queues;
    \item \convsystolicxqueue{}: systolic links implemented through \xqueue{}-managed queues;
    \item \convsystolicqlr{}: systolic links autonomously managed by the \gls{qlr} hardware.
\end{itemize}
With all \textit{conv2d} variants implementing the same functionality and systolic topology, the \gls{ipc} is stable around \SI{70}{\percent}. The utilization, however, largely depends on the implementation of the systolic links.
Software-emulated queues have a large overhead in terms of manual queue bookkeeping, which determines a high number of memory accesses and extremely low utilization.
Thanks to hardware-accelerated queue management, \convsystolicxqueue{} has \by{5}{} better utilization, with \texttt{q.push} and \texttt{q.pop} considerably reducing the number of control instructions.

As both \convsystolicsw{} and \convsystolicxqueue{} make use of explicit queue management, synchronization among \glspl{pe} happens through \gls{raw} stalls, which represent about 20\% of their runtime.
\glspl{qlr} fix this inconvenience by autonomously and implicitly managing the systolic communication among \glspl{pe}, reducing the control instructions to the bare minimum.
By handling queue synchronization through FIFOs, \glspl{qlr} also hide memory latencies and other \glspl{pe}' stalls. Overall, this further improves utilization of almost \by{2}{} (or \by{10}{} with respect to \convsystolicsw{}).
A minimum quantity of \gls{qlr} stalls is also visible for \convsystolicqlr{}, which depends on the depth of the \gls{qlr}'s FIFO.
As far as peak performance is concerned, \convsystolicxqueue{} and \convsystolicqlr{} are respectively \by{4}{} and \by{7}{} faster than \convsystolicsw{}, and respectively \by{3}{} and \by{6}{} more energy efficient.

While proving the concept of hybrid systolic-shared-memory architecture, the software-emulated queues are subject to massive control overhead.
While \xqueue{} represents a step closer to achieving efficient systolic execution, \convsystolicxqueue{} performance is undermined by the need for explicit queue accesses and limited support to in-flight queue operations, exclusively left to the processor's scoreboard. \gls{qlr} extension fully addresses these issues.
%
At a minimal area cost, \glspl{qlr} indeed allow \convsystolicqlr{} to achieve a steady-state utilization of about \SI{70}{\percent} and an energy efficiency of \SI{203}{\giga\ops\per\watt}.

\begin{table}[t!]
  \centering
  \caption{Evaluated hybrid implementations of the \emph{matmul} benchmark. Their incremental differences are highlighted in \textcolor{BrickRed}{red}.}
  \label{tab:systolic_matmul}
  \vspace{-0.2cm}
  \begin{tabular}{@{}lccccc@{}}
    \toprule
    Kernel & Tiling & \begin{tabular}[c]{@{}c@{}}Systolic\\array\end{tabular} & \begin{tabular}[c]{@{}c@{}}Tile\\mapping\end{tabular} &  \begin{tabular}[c]{@{}c@{}}Hybrid\\input load\end{tabular} &  \begin{tabular}[c]{@{}c@{}}Mover\\\glspl{pe}\end{tabular} \\
    \midrule
    \emph{matmul}\ensuremath{_{\text{\,QLR,\,1}}} & \by{2}{2} & \by{16}{16} & \by{2}{2} &  &  \\
    \emph{matmul}\ensuremath{_{\text{\,QLR,\,2}}} & \textcolor{BrickRed}{\by{3}{3}} & \by{16}{16} & \by{2}{2} &  &  \\
    \emph{matmul}\ensuremath{_{\text{\,QLR,\,3}}} & \textcolor{BrickRed}{\by{3}{4}} & \by{16}{16} & \by{2}{2} &  &  \\
    \emph{matmul}\ensuremath{_{\text{\,QLR,\,4}}} & \textcolor{BrickRed}{\by{3}{6}} & \by{16}{16} & \by{2}{2} &  &  \\
    \emph{matmul}\ensuremath{_{\text{\,QLR,\,5}}} & \textcolor{BrickRed}{\by{4}{4}} & \by{16}{16} & \by{2}{2} & \textcolor{BrickRed}{\checkmark} & \\
    \emph{matmul}\ensuremath{_{\text{\,QLR,\,6}}} & \by{4}{4} & \by{16}{16} & \by{2}{2} & \checkmark & \textcolor{BrickRed}{\checkmark} \\
    \emph{matmul}\ensuremath{_{\text{\,QLR,\,7}}} & \by{4}{4} & \by{16}{16} & \textcolor{BrickRed}{\by{1}{4}}    & \checkmark & \checkmark \\
    \emph{matmul}\ensuremath{_{\text{\,QLR,\,8}}} & \by{4}{4} & \textcolor{BrickRed}{\by{8}{32}}  & \by{1}{4}    & \checkmark & \checkmark \\
    \bottomrule
  \end{tabular}
\end{table}

\begin{figure}[t!]
    \centering
    \includegraphics{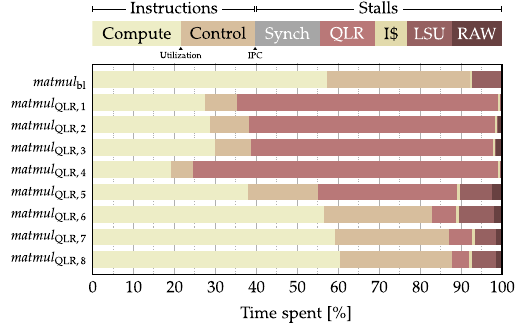}
    \vspace{-5.5mm}
    \caption{Relative performance results of the end-to-end \emph{matmul} benchmark, in its baseline (shared-memory) and systolic flavors.}
    \label{fig:stalls_matmul}
\end{figure}

\begin{figure}[t!]
    \centering
    \includegraphics[width=\columnwidth]{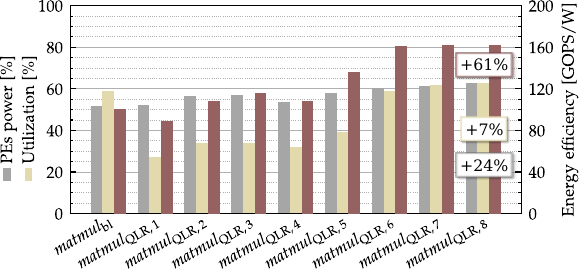}
    \vspace{-5mm}
    \caption{Steady-state analysis of the \emph{matmul} benchmark, in its baseline (shared-memory) and systolic flavors. The improvement with respect to \matmulbaseline{} is highlighted.}
    \label{fig:perfpower_matmul}
\end{figure}

Finally, we also evaluate the performance and energy efficiency of \convbaseline{} on both \mempoolbaseline{} and \mempoolqlr{}. In \mempoolqlr{}, when no \xqueue{} operation is issued and \glspl{qlr} are not configured, their related hardware is gated. This results in no performance or power penalties when systolic links are not stimulated, preventing any performance or power loss for general-purpose workloads.


\subsection{Evaluation of Hybrid Execution Models}
\label{subsec:eval_topology}

In this section, we evaluate the results of the \gls{dse} of \Cref{sec:kernels} for \emph{matmul} and \emph{conv2d}. For this analysis, we fix the hardware architecture to \mempoolqlr{}.

\subsubsection{Matrix Multiplication}

\Cref{tab:systolic_matmul} summarizes the hybrid variants of the \textit{matmul} benchmark evaluated in \Cref{fig:stalls_matmul} and \Cref{fig:perfpower_matmul}. 
Our evaluation starts from \matmulsystolicqlr{1}, which utilizes all four available \glspl{qlr} per core. With balanced vertical and horizontal systolic links, each \gls{pe} performs 4 \glspl{mac} for 8 queue operations\footnote{including both pushes to and pops from memory-mapped queues.}, resulting in a steady-state utilization of 27\%. Such a limited utilization is mainly determined by the low degree of data reuse: as the inner \glspl{pe} consume input operands too fast, the boundary \glspl{pe} cannot keep their pace in loading new ones, making \matmulsystolicqlr{1} memory-bound. This starves the downstream \glspl{pe}, determining \SI{65}{\percent} of the end-to-end runtime being spent in \gls{qlr} stalls.
With unbalanced vertical and horizontal systolic links, \matmulsystolicqlr{2} increases the data reuse of the input matrix elements, achieving 9 \glspl{mac} per 12 queue operations. 
In the same way, \matmulsystolicqlr{3} pushes it to 12 \glspl{mac} per 14 queue operations.
Exclusively with improved data reuse, \matmulsystolicqlr{3} achieves \SI{26}{\percent} higher utilization and \SI{31}{\percent} better energy efficiency with respect to the initial \matmulsystolicqlr{1}.
With a purely systolic execution model, further attempts to increase performance through higher data reuse fail, as demonstrated by \matmulsystolicqlr{4}: here, the inner \glspl{pe} become the bottleneck due to the highly unbalanced systolic links in the vertical direction.

However, a hybrid dataflow can still improve data reuse. With \matmulsystolicqlr{5}, each \gls{pe} performs 16 \glspl{mac} with only 8 queue operations and 4 load instructions. Despite the increment in load instructions, this approach increases the computational intensity of \emph{matmul} more efficiently than \matmulsystolicqlr{4}, bringing the steady-state utilization to 40\%.

\matmulsystolicqlr{5} achieves the upper bound regarding data reuse, determined by the maximum utilization of the register file. With their memory access overhead, the input boundary \glspl{pe} are now the bottleneck.
By relieving them of the computing tasks, \matmulsystolicqlr{6} achieves 50\% higher utilization. With horizontal systolic links only, \matmulsystolicqlr{7} minimizes the remote systolic links in the topology by mapping the tiles with \by{1}{4} grids of \glspl{pe}. This boosts the utilization of an additional 5\%. \matmulsystolicqlr{8} finally brings it to 63\% by reducing the number of systolic chains, and therefore of mover \glspl{pe}, from 16 to eight.
Overall, the optimal systolic topology of \matmulsystolicqlr{8} achieves a doubled steady-state utilization with respect to \matmulsystolicqlr{1}. The energy efficiency goes from \SI{89}{\giga\ops\per\watt} to \SI{163}{\giga\ops\per\watt}, also doubling.

\subsubsection{2D Convolution}

\begin{figure}[t!]
    \centering
    \includegraphics{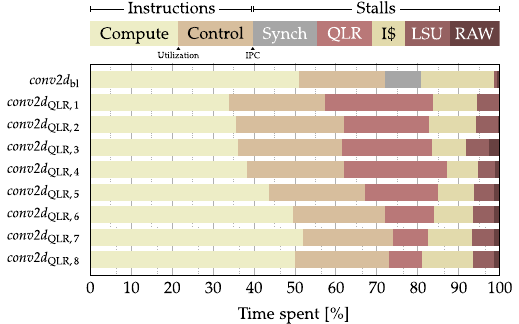}
    \vspace{-5.5mm}
    \caption{Relative performance results of the end-to-end \emph{conv2d} execution, in its baseline (shared-memory) and systolic flavors.}
    \label{fig:stalls_conv2d}
\end{figure}

\begin{figure}[t!]
    \centering
    \includegraphics[width=\columnwidth]{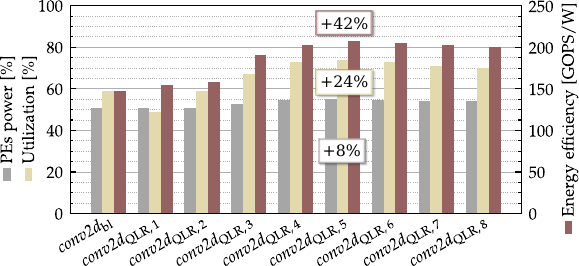}
    \vspace{-5mm}
    \caption{Steady-state analysis of the \emph{conv2d} benchmark, in its baseline (shared-memory) and systolic flavors. The improvement with respect to \convbaseline{} is highlighted.}
    \label{fig:perfpower_conv2d}
\end{figure}

\Cref{tab:systolic_conv2d} summarizes the hybrid variants of the \textit{conv2d} benchmark evaluated in \Cref{fig:stalls_conv2d} and \Cref{fig:perfpower_conv2d}. 
The starting point of our evaluation is \convsystolicqlr{1}. Thanks to the usage of the hybrid systolic-shared-memory execution model, it requires 4 queue operations and 1 load instruction for 9 \glspl{mac}. It achieves a steady-state utilization of 50\% and \SI{155}{\giga\ops\per\watt}. Only one long chain of \glspl{pe} is employed, which creates a large set-up time for the systolic array to fill up. This brings the utilization of the end-to-end kernel down to 34\%, with 26\% of the runtime composed of \glspl{qlr}-related stalls propagated through the long pipeline.
To balance the workload throughout the systolic array, \convsystolicqlr{2} relieves the chain-head \gls{pe} of computational tasks, increasing the steady-state utilization of 20\%.

\convsystolicqlr{3} and \convsystolicqlr{4} show the improvement exclusively related to data reuse optimization.
While \convsystolicqlr{3} requires 4 queue operations and 2 loads for 18 \glspl{mac}, \convsystolicqlr{4} only needs 4 queue operations and 3 loads for 27 \glspl{mac}. This results in an improvement of 23\% in steady-state utilization and 27\% in energy efficiency with respect to \convsystolicqlr{2}.
Increasing the degree of data reuse beyond this point is detrimental due to the limited capacity of the register file.
Therefore, with the remaining \textit{conv2d} implementations, we explore the optimum in terms of systolic chain length and, therefore, number of mover \glspl{pe}.

\begin{table}[t!]
  \centering
  \caption{Evaluated hybrid implementations of the \emph{conv2d} benchmark. Their incremental differences are highlighted in \textcolor{BrickRed}{red}.}
  \label{tab:systolic_conv2d}
  \vspace{-0.2cm}
  \begin{tabular}{@{}lccc@{}}
    \toprule
    Kernel & \begin{tabular}[c]{@{}c@{}}Input\\tiling\end{tabular} & \begin{tabular}[c]{@{}c@{}}Systolic\\array\end{tabular} & \begin{tabular}[c]{@{}c@{}}Mover\\\glspl{pe}\end{tabular} \\
    \midrule
    \emph{conv2d}\ensuremath{_{\text{\,QLR,\,1}}} & \by{3}{1} & 1 256-\gls{pe} chain &     \\
    \emph{conv2d}\ensuremath{_{\text{\,QLR,\,2}}} & \by{3}{1} & 1 256-\gls{pe} chain & \textcolor{BrickRed}{\checkmark} \\
    \emph{conv2d}\ensuremath{_{\text{\,QLR,\,3}}} & \textcolor{BrickRed}{\by{4}{1}} & 1 256-\gls{pe} chain & \checkmark \\
    \emph{conv2d}\ensuremath{_{\text{\,QLR,\,4}}} & \textcolor{BrickRed}{\by{5}{1}} & 1 256-\gls{pe} chain & \checkmark \\
    \emph{conv2d}\ensuremath{_{\text{\,QLR,\,5}}} & \by{5}{1} & \textcolor{BrickRed}{2 128-\gls{pe} chains} & \checkmark \\
    \emph{conv2d}\ensuremath{_{\text{\,QLR,\,6}}} & \by{5}{1} & \textcolor{BrickRed}{4 64-\gls{pe} chains} &  \checkmark \\
    \emph{conv2d}\ensuremath{_{\text{\,QLR,\,7}}} & \by{5}{1} & \textcolor{BrickRed}{8 32-\gls{pe} chains} &  \checkmark \\
    \emph{conv2d}\ensuremath{_{\text{\,QLR,\,8}}} & \by{5}{1} & \textcolor{BrickRed}{16 16-\gls{pe} chains} &  \checkmark \\
    \bottomrule
  \end{tabular}
\end{table}

As in \Cref{fig:perfpower_conv2d}, \convsystolicqlr{5} achieves the best steady-state utilization (\SI{74}{\percent}) and energy efficiency (\SI{208}{\giga\ops\per\watt}) with a topology composed of two systolic chains of 128 \glspl{pe} each. However, the highest utilization for the end-to-end benchmarks (\Cref{fig:stalls_conv2d}) is achieved by \convsystolicqlr{7} (\SI{52}{\percent}). Indeed, shorter chains imply shorter transient times but lower peak performance (\SI{71}{\percent} with \convsystolicqlr{7}'s eight chains), as less computing \glspl{pe} are available. A higher number of shorter, independent chains also positively impact the number of \gls{qlr} stalls, which stay confined to each individual chain. This is demonstrated by the reduction from 13.5\% to 6\% of the \gls{qlr} stalls proportion when moving from one (\convsystolicqlr{4}) to eight chains (\convsystolicqlr{7}).
However, based on the problem size, an optimal number of chains exists. A number of chains higher than the optimum jeopardizes the benefits from the faster transients as less compute \glspl{pe} are available (\convsystolicqlr{8}).


\subsection{Comparison with the Shared-Memory Architecture}
In this section, we evaluate the improvement of the hybrid systolic-shared-memory architecture over the baseline \mempool{} design. Specifically, we compare the baseline, highly-optimized benchmarks \matmulbaseline{}, \convbaseline{}, \cfftbaseline{}, and \dotpbaseline{} running on \mempoolbaseline{}, with their best hybrid implementations running on \mempoolqlr{}.

At steady state, \matmulsystolicqlr{8} has a utilization of 63\%, 7\% better than \matmulbaseline{}, and an energy efficiency of \SI{163}{\giga\ops\per\watt}, 61\% higher (\Cref{fig:perfpower_matmul}).
As measured in \cite{riedel2023mempool}, approximately 30\% of the power consumed at the \mempool{} group level when running \matmulbaseline{} is spent within the \gls{spm} interconnect. Moreover, it is proven that load operations from remote memory banks require, on average, double the energy of loads from the same tiles.
With respect to \matmulbaseline{}, \matmulsystolicqlr{8} spends \SI{7}{\percent} less of the end-to-end runtime in control instructions (\Cref{fig:stalls_matmul}), as most of the communication is implicitly handled by \glspl{qlr}. Moreover, the power consumption benefits from minimizing the number of remote systolic links.
This is demonstrated by the portion of the total power that is consumed in the \glspl{pe} for functional operations, which is \SI{52}{\percent} for \matmulbaseline{} and \SI{63}{\percent} for \matmulsystolicqlr{8}.

\rebuttal{
\convsystolicqlr{5} is less memory-intensive than the evaluated \textit{matmul} kernels, and makes a higher degree of data reuse possible at the level of the \glspl{pe}' register file. Therefore, the benefits of more efficient memory accesses have a smaller impact than what seen with \textit{matmul}, resulting in a energy efficiency of \SI{208}{\giga\ops\per\watt} (40\% higher than \convbaseline{}) The performance improvement, on the other hand, is even more remarkable, with a steady-state utilization of 73\% (24\% higher than \convbaseline{}).
}
\convsystolicqlr{5} also reports a portion of the total power spent in functional units of \SI{55}{\percent}, 4 percent points higher than \convbaseline{}.
Additionally, from the end-to-end kernel benchmarking (\Cref{fig:stalls_conv2d}), it can be observed how the \glspl{qlr} implicitly take care of \glspl{pe} synchronization, which makes explicit synchronization barriers (\SI{10}{\percent} of the \convbaseline{} runtime) of no use.

\begin{figure}[t!]
    \centering
    \includegraphics{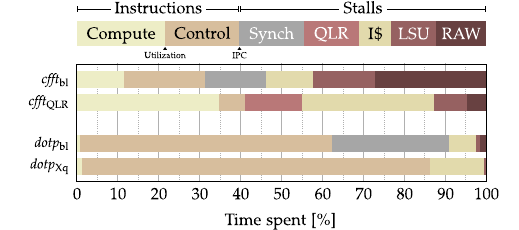}
    \vspace{-5.5mm}
    \caption{Relative performance results for the end-to-end \emph{cfft} and \textit{dotp} benchmarks, in their baseline (shared-memory) and systolic flavor. For \textit{dotp}, only the reduction-sum phase is considered.}
    \label{fig:stalls_cfft}
\end{figure}

\begin{figure}[t!]
    \centering
    \includegraphics[width=\columnwidth]{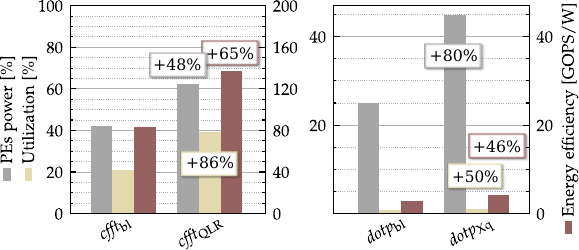}
    \vspace{-5mm}
    \caption{Steady-state analysis of the \emph{cfft} and \textit{dotp} benchmarks, in their baseline (shared-memory) and systolic flavor. The energy efficiency (right axis) is reported in 16-bit fixed-point operations for \textit{cfft}, and in 32-bit integer operations for \textit{dotp}. For \textit{dotp}, only the reduction-sum phase is considered.}
    \label{fig:perfpower_cfft}
\end{figure}

Due to its dataflow particularly suitable for systolic implementations, the radix-4 Cooley-Tukey \emph{cfft} benchmark is the one whose steady-state utilization benefits the most from the hybrid architecture. As a matter of fact, \cfftsystolicqlr{} achieves a utilization of \SI{39}{\percent} (\Cref{fig:perfpower_cfft}), almost doubling with respect to \cfftbaseline{} (\SI{21}{\percent}).
Although \cfftbaseline{} exploits both fine-grained and coarse-grained parallelization approaches~\cite{bertuletti2023efficient}, it has a large synchronization overhead (\Cref{fig:stalls_cfft}), which manifests itself as both sleep cycles (\SI{15}{\percent}) and \gls{raw} stalls (\SI{30}{\percent}).
On the other hand, \cfftsystolicqlr{}'s systolic topology allows each \gls{pe} to compute always the same \gls{fft} stage, processing the butterflies of different \glspl{fft} in a pipelined fashion. This cancels the need for global synchronization points, leaving the \glspl{qlr} only with the handling of the flow control required for synchronization among systolically linked \glspl{pe}.
The optimized communication topology and enhanced utilization increase the portion of total power consumed in the \glspl{pe} from \SI{42}{\percent} to \SI{62}{\percent}, bringing the energy efficiency of \cfftsystolicqlr{} to \SI{137}{\giga\ops\per\watt} (for 16-bit fixed-point datatype), \SI{65}{\percent} higher than \cfftbaseline{}.


\rebuttal{
As mentioned in \Cref{sec:kernels}, a hybrid systolic-shared-memory implementation of a kernel is beneficial as long as at least a minimal degree of communication and data reuse is possible among the \glspl{pe}. \Cref{fig:stalls_cfft} and \Cref{fig:perfpower_cfft} report the evaluation of \textit{dotp}'s reduction-sum phase. We do not consider \textit{dotp}'s first phase because, due to the absence of data reuse, its implementation is the same for both \mempoolbaseline{} and \mempoolqlr{}, and resorts to a purely shared-memory model. It is evaluated in \cite{riedel2023mempool}.
Our systolic implementation of the \textit{dotp} reduction-sum kernel, \dotpxqueue{}, doubles all figures of interest of \dotpbaseline{}. It is based on a radix-16 reduction tree implemented through \xqueue{}-managed systolic links. The functional unit utilization, measured in terms of 32-bit additions, reaches \SI{1.15}{\percent}, with an energy efficiency of \SI{4.3}{\giga\ops\per\watt}. When running \dotpxqueue{}, \SI{45}{\percent} of the power is spent in useful computation.
Similarly, to the \textit{cfft} case, the considerable improvement in the relative figures is mainly due to the absence of expensive synchronization points. \dotpbaseline{} implements the reduction tree through logarithmic barriers (\SI{30}{\percent} of its runtime), which force all cores to wait until the whole reduction is complete. On the other hand, \dotpxqueue{} exploits implicit synchronization through \xqueue{}-managed queues, that allow each core to independently terminate or run another workload after it completes its task in the reduction tree.

However, despite the entity of \dotpxqueue{}'s relative improvement, its absolute improvement is minimal. This is due to \textit{dotp}'s very limited data reuse. Moreover, the performance boost in the reduction phase becomes increasingly negligible with bigger input array sizes, due to the first phase dominating \textit{dotp}'s performance.
}


\section{Conclusions}
\label{sec:conclusions}

This work presents a hybrid systolic-shared-memory paradigm to enhance systolic computation in general-purpose, shared-L1-memory clusters. Our hybrid architecture demonstrates that L1-shared clusters with a low-latency interconnect can effectively implement systolic array topologies to boost the performance and energy efficiency of regular, compute-intensive workloads. Thanks to the low-latency interconnect, cores can act as \glspl{pe} of a systolic array, communicating in a user-defined, reconfigurable topology through queues mapped in the shared memory.
With the multicast and gather capabilities provided by the globally shared L1 memory, such an architecture can effectively combine the highly efficient systolic dataflow with the highly flexible concurrent memory accesses. By joining the best of both worlds, our hybrid architecture fills the gap between general-purpose, flexible architectures and specialized, highly efficient systolic arrays.

We propose two lightweight hardware extensions to support hybrid systolic-shared-memory computation, namely \xqueue{} and \glspl{qlr}. Accelerating memory-mapped queues and making them autonomous and implicit makes the hybrid architecture more efficient than a purely shared-memory cluster.
We develop an open-source reference implementation of the \xqueue{}- and \qlr{}-extended hybrid architecture based on \mempool{}~\cite{riedel2023mempool}, a 256-core \riscv{} manycore system with \SI{1}{\mega\byte} of L1 shared memory and a low-latency hierarchical interconnect.

We evaluate our hybrid architecture over a broad set of benchmarks based on four main \gls{dsp} kernels with diverse arithmetic intensity, namely \emph{matmul}, \emph{conv2d}, \emph{cfft}, and \emph{dotp}, exploring the vast trade-off space opened by the hybrid execution model.
When implemented for the GlobalFoundries' 22FDX \gls{fdsoi} \rebuttal{technology, in worst-case conditions (SS/\SI{0.72}{\volt}\kern-.15em/\SI{125}{\celsius}),} the \xqueue{} and \qlr{} extensions only impact \mempool{} with an area overhead of \SI{6}{\percent} at the level of the \mempool{} group (about \SI{11}{\kilo\ge} per core), without causing any frequency degradation. At such a cost, \glspl{qlr} make systolic workloads up to \by{7}{} faster and \by{6}{} more energy efficient than software-emulated queues, actually enabling efficient hybrid computation.

%
With high-arithmetic-intensity kernels, such as \emph{conv2d}, our hybrid architecture reaches a utilization of up to \SI{73}{\percent}. In our experiments, the hybrid architecture improves the utilization of compute-intensive kernels up to almost \by{2} with respect to the shared-memory baseline.
Moreover, with a hybrid execution model, we report a portion of total power spent in the \glspl{pe} up to \SI{63}{\percent}.
At \SI{600}{\mega\hertz} and in typical operating conditions (TT/\SI{0.80}{\volt}\kern-.16em/\SI{25}{\celsius}), the hybrid architecture achieves up to \SI{208}{\giga\ops\per\watt} for 32-bit operations, and reaches up to \SI{65}{\percent} improvement with respect to the baseline. These improvements come without any performance or energy efficiency degradation for kernels that cannot resort to hybrid systolic-shared-memory execution models, such as low-arithmetic-intensity, irregular workloads.


\Urlmuskip=0mu plus 1mu\relax
\def\UrlBreaks{\do\/\do-}
\bibliographystyle{IEEEtran}
\bibliography{main}

\end{document}

%% file: preamble.tex
\ifCLASSOPTIONcompsoc%
  \usepackage[nocompress]{cite}
\else
  \usepackage{cite}
\fi

\ifCLASSINFOpdf%
  \usepackage[pdftex]{graphicx}
  \graphicspath{{../pdf/}{../jpeg/}}
  \DeclareGraphicsExtensions{.pdf,.jpeg,.png}
\else
  \usepackage[dvips]{graphicx}
  \graphicspath{{../eps/}}
  \DeclareGraphicsExtensions{.eps}
\fi
\usepackage{mwe}
\usepackage[pdftex,dvipsnames]{xcolor}
\usepackage{xprintlen}
\usepackage{pgfplots}
\pgfplotsset{compat=1.16}
\usepackage{booktabs}
\usepackage{multirow}
\usepackage{tabularx}
\usepackage{adjustbox}
\usepackage{makecell}
\usepackage{rotating}
\usepackage{calc}
\usepackage{amsmath,amssymb,amsfonts}
\usepackage{algorithmic}
\usepackage[free-standing-units,per-mode=repeated-symbol,mode=text,detect-weight=true,detect-family=true]{siunitx}
\usepackage{textcomp}
\usepackage{comment}

\usepackage{xargs}                      
\usepackage[colorinlistoftodos,prependcaption,textsize=tiny]{todonotes}
\newcommandx{\change}[2][1=]{\todo[linecolor=red,backgroundcolor=red!25,bordercolor=red,#1]{#2}}
\newcommandx{\info}[2][1=]{\todo[linecolor=OliveGreen,backgroundcolor=OliveGreen!25,bordercolor=OliveGreen,#1]{#2}}
\newcommandx{\unsure}[2][1=]{\todo[linecolor=Plum,backgroundcolor=Plum!25,bordercolor=Plum,#1]{#2}}

\usepackage{array}
\usepackage{parcolumns}
\usepackage{xspace}
\usepackage{listings}
\usepackage{enumitem}

\usepackage{url}
\usepackage{glossaries}
\usepackage[hidelinks]{hyperref}
\hypersetup{breaklinks=true}
\usepackage[capitalise,nameinlink]{cleveref}

\usepackage{microtype}
\usepackage{soul}

\definecolor{color1}{HTML}{f94144}
\definecolor{color2}{HTML}{f9c74f}
\definecolor{color3}{HTML}{90be6d}
\definecolor{color4}{HTML}{577590}
\colorlet{colorAlert}{Red}

\newcounter{undefinedreferences}
\newcommand{\checkreferences}{
	\ifnum\value{undefinedreferences} > 0
	\begin{center}
		\immediate\write18{wget -O Figures/protester.png -nc http://imgs.xkcd.com/comics/wikipedian_protester.png}
		\includegraphics[width=\textwidth]{protester.png}
	\end{center}
	\else
	No undefined references. Good!
	\fi
}

\DeclareSIUnit\core{core}
\DeclareSIUnit\tile{tile}
\DeclareSIUnit\request{request}
\DeclareSIUnit\cycle{cycle}
\DeclareSIUnit\mac{MACs}
\DeclareSIUnit\erlang{E}
\DeclareSIUnit\flop{FLOP}
\DeclareSIUnit\flops{FLOPS}
\DeclareSIUnit\gate{GE}
\DeclareSIUnit\ge{GE}
\DeclareSIUnit\op{OP}
\DeclareSIUnit\ops{OPS}
\DeclareSIUnit\bps{bps}
\DeclareSIUnit\Bps{Bps}
\DeclareSIUnit\ipc{IPC}
\DeclareSIUnit\byte{B}
\DeclareSIUnit\ips{IPS}
\DeclareSIUnit[number-unit-product = ]\percent{\%}

\PackageWarning{Citation missing:}{#1!}

\PackageWarning{TODO:}{#1!}

\providecommand{\rebuttal}[1]{{#1}}

\newcommand\etal{et\penalty50\ al.\ }
\newcommand\riscv{RISC-V}
\newcommand\mempool{Mem\-Pool}
\newcommand\qpush{\texttt{queue\_push}}
\newcommand\qpop{\texttt{queue\_pop}}

\newcommand\xqueue{Xqueue}
\newcommand\qlr{QLR}
\newcommand{\SIadj}[2]{\SI[number-unit-product={\text{-}}]{#1}{#2}}
\newcommand\by[2]{#1\ensuremath{\times}#2}
\newcommand\benchmark[3]{\textit{#1}\ifx&#3&\ensuremath{_{\text{\,#2}}}\else\ensuremath{_{\text{\,#2,\,#3}}}\fi}

\newcommand\matmul{\emph{matmul}}

%% file: acronyms.tex
\newacronym{udma}{\textmu DMA}{micro direct memory access}
\newacronym{abi}{ABI}{application binary interface}
\newacronym{ace}{ACE}{AXI Coherent Extensions}
\newacronym{alu}{ALU}{arithmetic logic unit}
\newacronym{amba}{AMBA}{Advanced Microcontroller Bus Architecture}
\newacronym{amo}{AMO}{atomic memory operation}
\newacronym{aot}{AOT}{ahead-of-time}
\newacronym{apb}{APB}{Advanced Peripheral Bus}
\newacronym{api}{API}{application programming interface}
\newacronym{asic}{ASIC}{application-specific integrated circuit}
\newacronym{axi}{AXI}{Advanced eXtensible Interface}
\newacronym{blas}{BLAS}{Basic Linear Algebra Subprograms}
\newacronym{cas}{CAS}{compare-and-swap}
\newacronym[longplural={core complexes}]{cc}{CC}{core complex}
\newacronym{cgra}{CGRA}{coarse-grained reconfigurable architecture}
\newacronym{cmos}{CMOS}{complementary metal-oxide-semiconductor}
\newacronym{cnn}{CNN}{convolutional neural network}
\newacronym{cpu}{CPU}{central processing unit}
\newacronym{csr}{CSR}{control and status register}
\newacronym{dbt}{DBT}{dynamic binary translation}
\newacronym{dct}{DCT}{discrete cosine transform}
\newacronym{dit_adj}{DIT}{decimation-in-time}
\newacronym{dif_adj}{DIF}{decimation-in-frequency}
\newacronym{dlp}{DLP}{data-level parallelism}
\newacronym{dma}{DMA}{direct memory access}
\newacronym{dram}{DRAM}{dynamic random-access memory}
\newacronym{dse}{DSE}{design space exploration}
\newacronym{dsl}{DSL}{domain-specific language}
\newacronym{dsp}{DSP}{digital signal processing}
\newacronym{elf}{ELF}{Executable and Linkable Format}
\newacronym{fdsoi}{FD-SOI}{fully depleted silicon-on-insulator}
\newacronym{fft}{FFT}{Fast Fourier transform}
\newacronym{dft}{DFT}{Discrete Fourier transform}
\newacronym{fifo}{FIFO}{first in, first out}
\newacronym{fpga}{FPGA}{field-programmable gate array}
\newacronym{fpu}{FPU}{floating-point unit}
\newacronym{fsm}{FSM}{finite-state machine}
\newacronym{gemm}{GEMM}{general matrix multiplication}
\newacronym{gpgpu}{GPGPU}{general-purpose computing on a \gls{gpu}}
\newacronym{gpu}{GPU}{graphics processing unit}
\newacronym{hart}{hart}{hardware thread}
\newacronym{hbm}{HBM}{High Bandwidth Memory}
\newacronym{hdl}{HDL}{hardware description language}
\newacronym{hero}{HERO}{Heterogeneous Embedded Research Platform}
\newacronym{hpc}{HPC}{high-performance computing}
\newacronym{ilp}{ILP}{instruction level parallelism}
\newacronym{ipc}{IPC}{instructions per cycle}
\newacronym{iot}{IoT}{Internet of Things}
\newacronym{ipu}{IPU}{integer processing unit}
\newacronym{ir}{IR}{intermediate representation}
\newacronym{isa}{ISA}{instruction set architecture}
\newacronym{issr}{ISSR}{indirection stream semantic register}
\newacronym{jit}{JIT}{just-in-time}
\newacronym{llc}{LLC}{last-level cache}
\newacronym{lrsc}{LRSC}{load-reserved/store-conditional}
\newacronym{lr}{LR}{load-reserved}
\newacronym{lsu}{LSU}{load-store unit}
\newacronym{mac}{MAC}{multiply–accumulate}
\newacronym{mimd}{MIMD}{multiple instruction, multiple data}
\newacronym{mmu}{MMU}{memory management unit}
\newacronym[longplural={networks-on-chip}]{noc}{NoC}{network-on-chip}
\newacronym{numa}{NUMA}{non-uniform memory access}
\newacronym{ppa}{PPA}{power, performance, and area}
\newacronym{pc}{PC}{program counter}
\newacronym{pe}{PE}{processing element}
\newacronym{pl}{PL}{programmable logic}
\newacronym{pmca}{PMCA}{programmable manycore accelerator}
\newacronym{pulp}{PULP}{Parallel Ultra Low Power}
\newacronym{qlr}{QLR}{queue-linked register}
\newacronym{raw}{RAW}{read-after-write}
\newacronym{rom}{ROM}{read-only memory}
\newacronym{rmw}{RMW}{read–modify–write}
\newacronym{rob}{ROB}{reorder buffer}
\newacronym{ro}{RO}{read-only}
\newacronym{rtl}{RTL}{register-transfer level}
\newacronym{sbt}{SBT}{static binary translation}
\newacronym{scm}{SCM}{standard cell memory}
\newacronym{sc}{SC}{store-conditional}
\newacronym{sdf}{SDF}{Standard Delay Format}
\newacronym{simd}{SIMD}{single instruction, multiple data}
\newacronym{simt}{SIMT}{single instruction, multiple thread}
\newacronym{sm}{SM}{streaming multiprocessor}
\newacronym{soc}{SoC}{system-on-chip}
\newacronym[longplural={scratchpad memories}]{spm}{SPM}{scratchpad memory}
\newacronym{sram}{SRAM}{static random-access memory}
\newacronym{ssa}{SSA}{static single assignment}
\newacronym{ssr}{SSR}{stream semantic register}
\newacronym{tcdm}{TCDM}{tightly-coupled data memory}
\newacronym{tlp}{TLP}{thread-level parallelism}
\newacronym{tpu}{TPU}{Tensor Processing Unit}
\newacronym{vpu}{VPU}{vector processing unit}
\newacronym{vliw}{VLIW}{very long instruction word}
\newacronym{vnb}{VNB}{von Neumann bottleneck}
\newacronym{war}{WAR}{write-after-read}
\newacronym{waw}{WAW}{write-after-write}